\documentclass[prc,aps,amsfonts,dvips,twocolumn,floatfix]{revtex4}
\usepackage{graphicx,amssymb,color}
\usepackage{epsf,epsfig,rotating,pgf,array,xspace}
\usepackage{ulem}

\definecolor{orange}{cmyk}{0,0.61,0.87,0}

\begin{document}

\title{\bf
Two-particle spatial correlations in superfluid nuclei}
\author{\rm N. Pillet$^{a}$,
N. Sandulescu$^{b}$,
P. Schuck$^{c,d,e}$,
J.-F. Berger$^{a}$}

\bigskip

\affiliation{\rm
$^{(a)}$ CEA/DAM/DIF, F-91297 Arpajon, France  \\
$^{(b)}$ Institute of Physics and Nuclear Engineering, 76900 Bucharest,
Romania \\
$^{(c)}$~  Institut de Physique Nucl\'eaire, CNRS, UMR8608, Orsay,
F-91406, France \\
$^{(d)}$~  Universit\'e Paris-Sud, Orsay,
F-91505, France \\
$^{(e)}$~ Laboratoire de Physique et Mod\'elisation des Milieux Condens\'es,
CNRS and Universit\'e Joseph Fourier, Maison des Magist\`eres,
Bo\^ite Postale 166, 38042 Grenoble Cedex, France}

\date{\today}

\def\fid{\vert\phi >}
\def\fig{< \phi\vert}
\def\psid{\vert\Psi>}
\def\psig{<\Psi\vert}
\def\psid{\vert\Psi>}
\def\psig{<\Psi\vert}
\def\dspt{\displaystyle}

\begin{abstract}
We discuss the effect of pairing on two-neutron space correlations in
deformed nuclei.
The spatial correlations are described by the pairing tensor in coordinate
space calculated in the HFB approach. The calculations are done using the
D1S Gogny force. 
We show that the pairing tensor has a rather small extension in
the relative coordinate, a feature observed earlier in spherical nuclei.
It is pointed out that in deformed nuclei the coherence length
corresponding to the pairing tensor has a pattern similar to what we have
found previously in spherical nuclei, i.e., it is maximal in the interior of
the nucleus and then it is decreasing rather fast in the surface region where
it reaches a minimal value of about 2 fm. This minimal value of the coherence 
length in the surface is essentially determined by the
finite size properties of single-particle states in the vicinity of
the chemical potential and has little to do with enhanced 
pairing correlations in the nuclear surface. It is shown that in nuclei 
the coherence length is not a good indicator of 
the intensity of pairing correlations. 
This feature is contrasted with the situation in infinite matter.
\end{abstract}

\maketitle

\section{Introduction}
According to pairing models, in open shell nuclei the nucleons with energies close
to the Fermi level form correlated Cooper pairs. One of the most obvious manifestation
of correlated pairs in nuclei is the large cross section for two particle
transfer.
In the HFB approach, commonly employed to treat pairing in nuclei, the pair transfer
amplitude is approximated by the pairing tensor. In coordinate space the pairing
tensor for like nucleons is defined by

\begin{equation}
\kappa \left( \vec{r_{1}} s_{1}, \vec{r_{2}} s_{2} \right) =
\langle HFB \vert  \Psi \left( \vec{r_{1}} s_{1} \right) \Psi
\left( \vec{r_{2}} s_{2} \right)
\vert HFB \rangle
\label{eq29}
\end{equation}
where $\vert HFB \rangle$ is the HFB ground state wave function
while $\Psi \left( \vec{r} s \right)$ is the nucleon field operator.
By definition, the pairing tensor
$\kappa \left( \vec{r_{1}} s_{1}, \vec{r_{2}} s_{2} \right)$
is the probability amplitude to find in the ground state of the system two correlated
nucleons with the positions $\vec{r_{1}}$ and
$\vec{r_{2}}$ and with the spins $s_1$ and $s_2$. This is the non-trivial
part of the two-body correlations which is not contained in the Hartree-Fock approximation.

In spite of many HFB calculations done for about half a century, there are only few
studies dedicated to the non-local spatial properties  of pairing tensor in
atomic nuclei \cite{tischler,matsuo,ref1,pastore}. One of the most interesting properties
of the pairing tensor revealed recently  is its small extension in the relative coordinate
$\vec{r}=\vec{r_1}-\vec{r_2}$. Thus in Ref. \cite{ref1} it is shown
that the averaged relative distance, commonly called the coherence length, has an unexpected small
value in the surface of spherical nuclei, of about 2-3 fm.  This value is about two  times smaller
than the lowest coherence length in infinite matter. Similar small values of coherence length have
been  obtained later for some spherical nuclei \cite{pastore} and for a slab of non-uniform
neutron matter \cite{pankratov}. \\
The scope of this paper is to extend the study done in Ref. \cite{ref1} and to investigate
axially-deformed nuclei. It will be shown that in axially-deformed nuclei the pairing tensor has
similar spatial features as in spherical nuclei, including a small coherence length in the nuclear
surface. The paper is organized as follows. In section \ref{sect1}, the general expression of the
pairing tensor is derived in an axially deformed harmonic oscillator basis. Expressions of the pairing
tensor coupled to a total spin S=0 or 1 and associated projection are also presented in three particular
geometrical configurations. In section \ref{sect2}, local as well as non-local part of the pairing tensor
are discussed for few axially deformed nuclei, namely $^{152}Sm$, $^{102}Sr$ and $^{238}U$. Results
concerning coherence length are also presented and interpreted in a less exclusive way compared
to Ref. \cite{ref1}. Summary and conclusions are given in section \ref{sect3}.

\section{Pairing tensor for axially-deformed nuclei} \label{sect1}

As in Ref.\cite{ref1}, we calculate the pairing tensor in the HFB approach
using the D1S Gogny force \cite{d1}. To describe axially-deformed nuclei
we take a single-particle
basis formed by axially-deformed harmonic oscillator (HO) wave functions.
In this basis the nucleon field operators can be written as

\begin{equation}
\Psi\left( \vec{r}, s \right) = \sum_{m \nu} c^{+}_{ms \nu} \phi_{m \nu}
\left( \vec{r} \right)
\label{eq19}
\end{equation}
where the HO wave function is
\begin{equation}
\phi_{m \nu} \left( \vec{r} \right) = e^{im \theta} ~\Re_{\vert m \vert \nu}
\left( \widetilde{r} \right)
\end{equation}
The quantum numbers $m$ and $s$ are the projection of the orbital and spin momenta on symmetry
(z) axis; $\nu$ are the radial quantum numbers $\nu= \left( n_{\bot}, n_{z}\right)$.
The function  $\Re_{\vert m \vert \nu} \left( \widetilde{r} \right) \equiv \Re_{\vert m \vert \nu}
\left( r_{\bot}, z \right)$ is given by
\begin{equation}
\Re_{\vert m \vert \nu} \left( r_{\bot}, z \right) = \varphi_{n_{z}} \left( z, \alpha_{z} \right)
\times \varphi_{n_{\bot} m} \left( r_{\bot}, \alpha_{\bot} \right)
\label{eq21}
\end{equation}
where
\begin{equation}
\varphi_{n_{z}} \left( z, \alpha_{z} \right) = \left( \frac{\alpha_{z}}{\pi} \right) ^{\frac{1}{4}}
\left[ \frac{1}{2^{n_{z}} n_{Z}!} \right]^{1/2} e^{\frac{1}{2} \alpha_{z} z^{2}} H_{n_{z}}
\left( z \sqrt{\alpha_{z}} \right)
\label{eq22}
\end{equation}
and
\begin{equation}
\begin{array}{c}
\varphi_{n_{\bot} \vert m \vert} \left( r_{\bot}, \alpha_{\bot} \right) =
\left( \frac{\alpha_{\bot}}{\pi} \right)^{1/2}
\left[ \frac{n_{\bot}!}{\left( n_{\bot} + \vert m \vert \right)!} \right]^{1/2} \\
\times ~e^{\frac{1}{2} \alpha_{\bot} r^{2}_{\bot}} \left( r_{\bot}
\sqrt{\alpha_{\bot}} \right)^{\vert m \vert}
L_{n_{\bot}}^{\vert m \vert} \left( \alpha_{\bot} r_{\bot}^{2} \right)
\end{array}
\end{equation}
In the above equations,  $\alpha_z$ and $\alpha_\perp$ are the HO parameters
in the $z$ and perpendicular directions, which are related to the HO
frequencies by $\alpha_z=M\omega_z/\hbar$ and
$\alpha_\perp=M\omega_\perp/\hbar$, respectively with $M$ the nucleon
mass, and $H_{n_{z}}$ and $L_{n_{\bot}}$ are Hermite and Laguerre
polynomials, respectively.

Using the expansion (2) it can be shown that the pairing tensor in coordinate
representation can be written in the following form (the spin up is donoted by
"+" and the spin down by "-")
\begin{equation}
\begin{array}{l}
\dspt \kappa \left( \vec{r_{1}} +, \vec{r_{2}} - \right) = \dspt
\sum_{m_{1} \ge 0~ \nu_{1} \nu_{2}}
~\Re_{\vert m_{1} \vert \nu_{1}} \left( \widetilde{r_{1}} \right)
~\Re_{\vert m_{1} \vert \nu_{2}} \left( \widetilde{r_{2}} \right) \\
\hspace{10mm}
\left( e^{im_{1}( \theta_{1} -\theta_{2}) } \,\,
\widetilde{\kappa}^{m_{1}+1/2}_{m_{1} \nu_{1}, m_{1} \nu_{2}} \right.
\\   \hspace{12mm}
+ \left.\left( 1- \delta_{m_{1},0} \right) e^{-im_{1} \left( \theta_{1} -
\theta_{2} \right) } \,\,
\widetilde{\kappa}^{m_{1}-1/2}_{m_{1} \nu_{1}, m_{1} \nu_{2}} \right)
\end{array}
\label{eq40}
\end{equation}
\begin{equation}
\begin{array}{l}
\dspt \kappa \left( \vec{r_{1}} +, \vec{r_{2}} + \right) =
-  \sum_{m_{1} \ge 0~ \nu_{1} \nu_{2}} \\ \dspt \hspace{5mm}
\left( e^{im_{1} \theta_{1}- i \left( m_{1} +1\right) \theta_{2}}
~\Re_{\vert m_{1} \vert \nu_{1}} \left( \widetilde{r_{1}} \right)
~\Re_{\vert m_{1}+1 \vert \nu_{2}} \left( \widetilde{r_{2}} \right)\right.
\\\dspt\hspace{2mm} \left.
-e^{-i\left( m_{1}+1 \right) \theta_{1}+ i m_{1} \theta_{2}}
~\Re_{\vert m_{1} \vert \nu_{1}} \left( \widetilde{r_{2}} \right)
~\Re_{\vert m_{1}+1 \vert \nu_{2}} \left( \widetilde{r_{1}} \right) \right)
\\\dspt\hspace{5mm}
\times \widetilde{\kappa}^{m_{1}+1/2}_{m_{1} \nu_{1},m_{1}+1 \nu_{2}}
\end{array}
\label{eq41}
\end{equation}

\begin{figure}
\begin{center}
\includegraphics*[scale=0.34,angle=0]{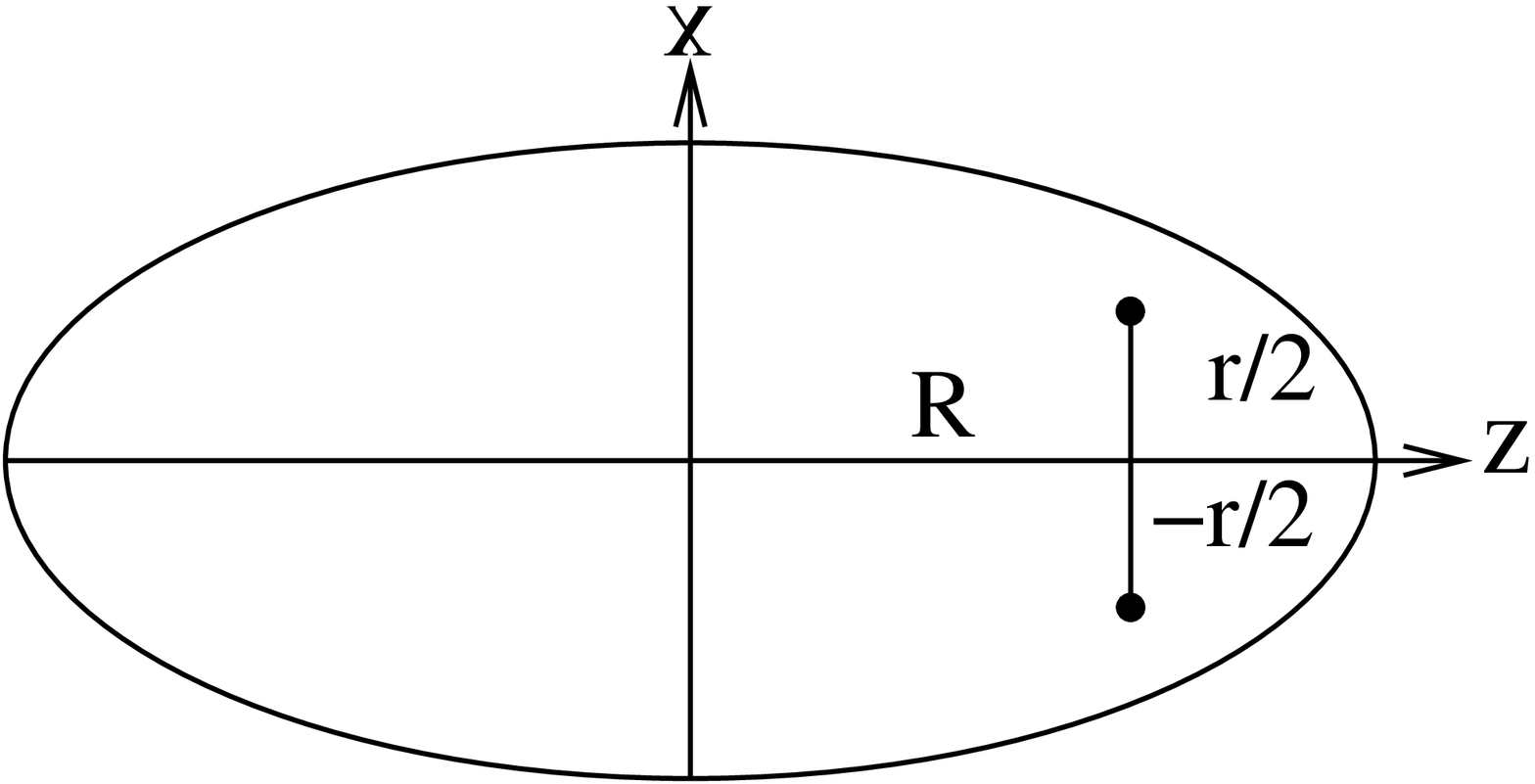}
\caption{ The geometrical configuration (a) corresponding
to two neutrons in the xz plane. R and r indicate the c.o.m
position and the relative distance of the two neutrons.}
\label{fig1}
\end{center}
\end{figure}
\begin{figure}
\begin{center}
\includegraphics*[scale=0.34,angle=0]{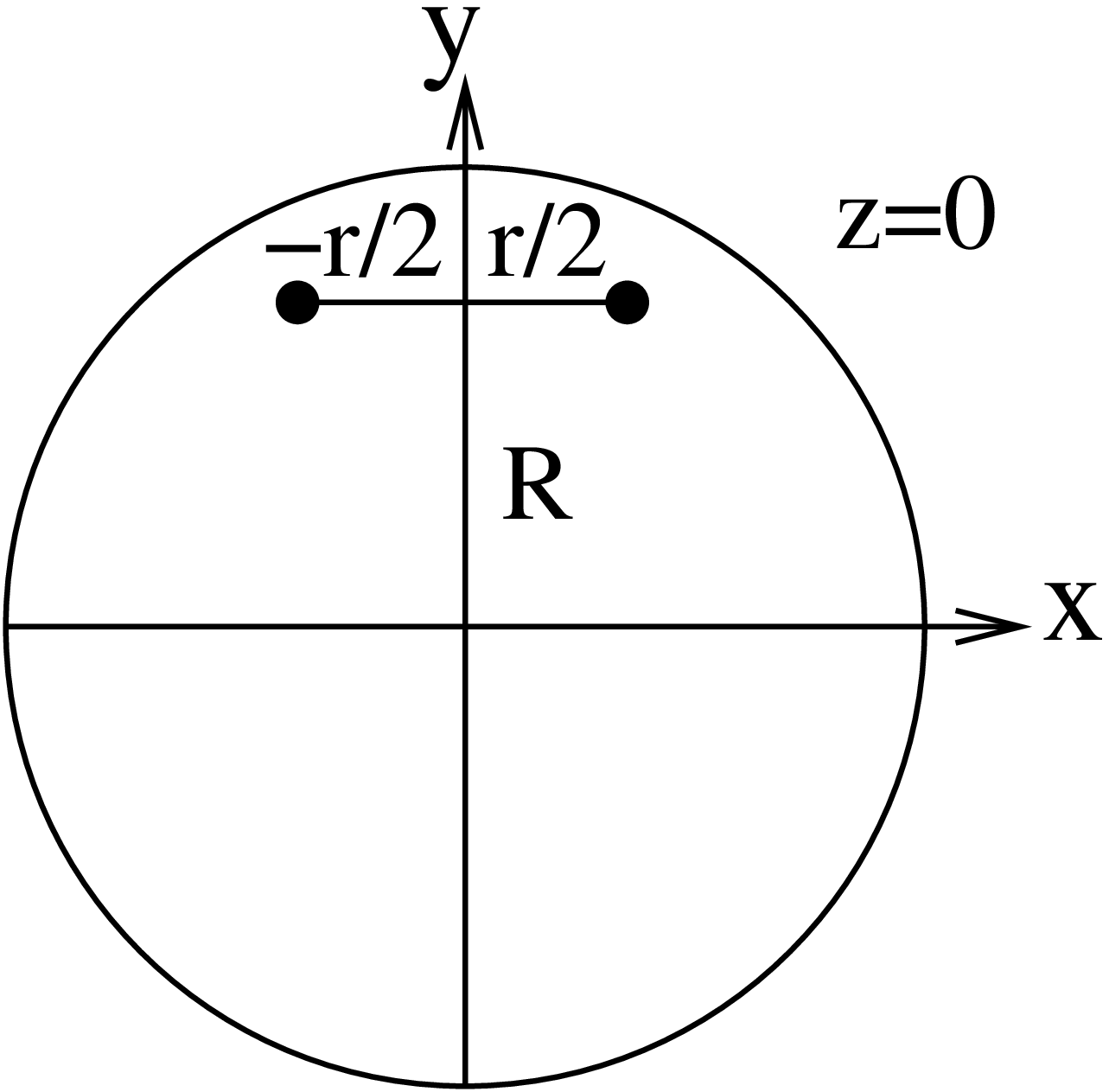}
\caption{ The geometrical configuration (b) corresponding to
two neutrons in the xy plane. R and r indicate the c.o.m
position and the relative distance of the two neutrons.}
\label{fig2}
\end{center}
\end{figure}
\begin{figure}
\begin{center}
\includegraphics*[scale=0.34,angle=0]{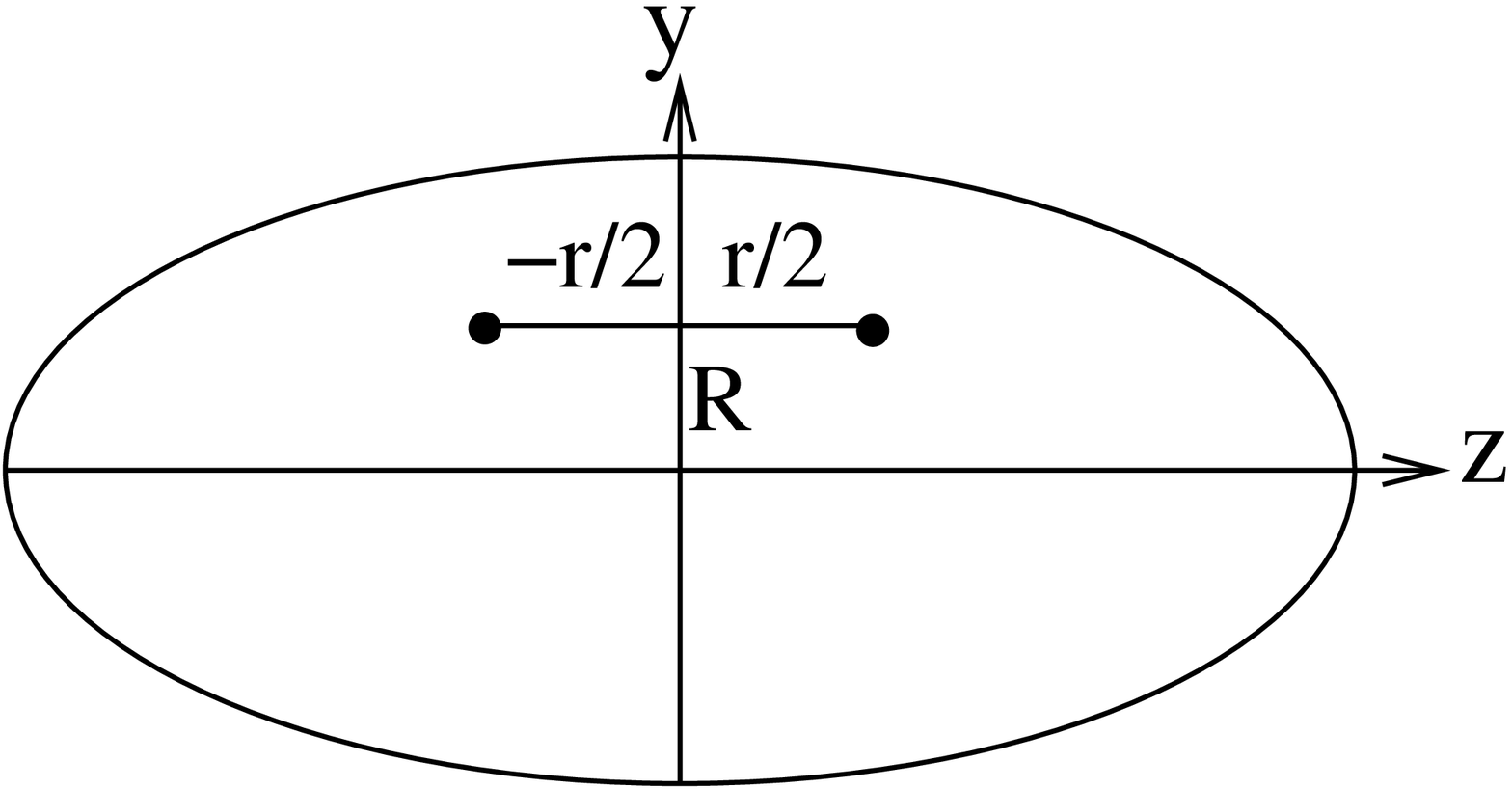}
\caption{ The geometrical configuration (c) corresponding to two
neutrons in the yz plane. R and r indicate the c.o.m
position and the relative distance of the two neutrons.}
\label{fig3}
\end{center}
\end{figure}

In the above expressions we have introduced the pairing tensor in the HO basis
\begin{equation}
\begin{array}{l}
\widetilde{\kappa}_{\alpha_{1} \alpha_{2}} \equiv
\widetilde{\kappa}^{\Omega}_{m_{1} \nu_{1}, m_{2} \nu_{2}} =
2s_{2} \langle \widetilde{0} \vert c_{m_{1} s_{1} \nu_{1}} c_{-m_{2} -s_{2}
\nu_{2}} \vert \widetilde{0} \rangle \\
\hspace{3.0cm} = \widetilde{\kappa}_{\alpha_{2} \alpha_{1}}
\end{array}
\label{eq36}
\end{equation}
where $\Omega=m_{1}+s_{1}=m_{2}+s_{2}$.

In the present study we calculate the pairing tensor corresponding to three geometrical
configurations shown in Figs.\ref{fig1}-\ref{fig3}; they have the advantage of a simple
separation between the center of mass (c.o.m) $\vec{R}=(\vec{r_1}+\vec{r_2})/2$
and the relative $\vec{r}=\vec{r_1}-\vec{r_2}$ coordinates.

For a finite range force, as the D1S Gogny force used here,
the pairing tensor has non-zero values for the total
spin $S=0$ and $S=1$. How these two channels are related to the pairing
tensors (\ref{eq40})-(\ref{eq41}) depends
on the geometrical configuration. Thus it can be shown that for the configuration displayed in
Fig.\ref{fig1} the following relations are satisfied:
\begin{equation}
\left[ \kappa \left( \vec{r_{1}} s_{1}, \vec{r_{2}} s_{2} \right) \right]_{00} =
\sqrt{2} ~\kappa \left( \vec{r_{1}} +, \vec{r_{2}} - \right)
\label{eq45}
\end{equation}
\begin{equation}
\left[ \kappa \left( \vec{r_{1}} s_{1}, \vec{r_{2}} s_{2} \right) \right]_{10} = 0
\label{eq46}
\end{equation}
\begin{equation}
\left[ \kappa \left( \vec{r_{1}} s_{1}, \vec{r_{2}} s_{2} \right) \right]_{11} =
\left[ \kappa \left( \vec{r_{1}} s_{1}, \vec{r_{2}} s_{2} \right) \right]_{1-1} =
\kappa \left( \vec{r_{1}} +, \vec{r_{2}} + \right)
\label{eq47}
\end{equation}
where the notation $[..]_{ij}$ means that the pairing tensor is coupled to total spin $S=i$
with the projection $S_{z}=j$.

For the configuration shown in Fig.\ref{fig2} the pairing tensor $\kappa \left( \vec{r_{1}}
s_1, \vec{r_{2}} s_2 \right)$ is a complex quantity and we have the relations
\begin{equation}
\left[ \kappa \left( \vec{r_{1}} s_{1}, \vec{r_{2}} s_{2} \right) \right]_{00} =
\sqrt{2} ~Re \left( \kappa \left( \vec{r_{1}} +, \vec{r_{2}} - \right) \right)
\label{eq51}
\end{equation}
\begin{equation}
\left[ \kappa \left( \vec{r_{1}} s_{1}, \vec{r_{2}} s_{2} \right) \right]_{10} =
i\sqrt{2} ~Im \left( \kappa \left( \vec{r_{1}} +, \vec{r_{2}} - \right) \right)
\label{eq52}
\end{equation}
\begin{equation}
\left[ \kappa \left( \vec{r_{1}} s_{1}, \vec{r_{2}} s_{2} \right) \right]_{11} =
\left[ \kappa \left( \vec{r_{1}} s_{1}, \vec{r_{2}} s_{2} \right) \right]_{1-1} =
\kappa \left( \vec{r_{1}} +, \vec{r_{2}} + \right)
\label{eq53}
\end{equation}
Finally, for the configuration (c) of Fig.\ref{fig3}, we have
\begin{equation}
\left[ \kappa \left( \vec{r_{1}} s_{1}, \vec{r_{2}} s_{2} \right) \right]_{00} =
\sqrt{2} ~Re \left( \kappa \left( \vec{r_{1}} +, \vec{r_{2}} - \right) \right)
\label{eq57}
\end{equation}
\begin{equation}
\left[ \kappa \left( \vec{r_{1}} s_{1}, \vec{r_{2}} s_{2} \right) \right]_{10} = 0
\label{eq58}
\end{equation}
\begin{equation}
\left[ \kappa \left( \vec{r_{1}} s_{1}, \vec{r_{2}} s_{2} \right) \right]_{11} =
\kappa \left( \vec{r_{1}} +, \vec{r_{2}} + \right)
\label{eq59}
\end{equation}
\begin{equation}
\left[ \kappa \left( \vec{r_{1}} s_{1}, \vec{r_{2}} s_{2} \right) \right]_{1-1} =
\kappa \left( \vec{r_{1}} -, \vec{r_{2}} - \right) =
\kappa^{*} \left( \vec{r_{1}} +, \vec{r_{2}} + \right)
\label{eq60}
\end{equation}
The results for the pairing tensor shown in this paper are obtained by solving the HFB equations 
in a HO basis with 13 major shells for deformed nuclei. We have checked that by increasing the dimension of
the basis the spatial properties of the pairing tensor do not change significantly up to distances of
about 10 fm in the nuclei studied here. This shows that a finite discrete 13 major shell HO basis
correctly describes these nuclei in the domain of interest of this work, in particular that
continuum coupling effects can be ignored.

\section{Results and Discussion} \label{sect2}

\subsection{Local and non-local parts of the pairing tensor}

\begin{figure}
\begin{center}
\includegraphics*[scale=0.45,angle=0]{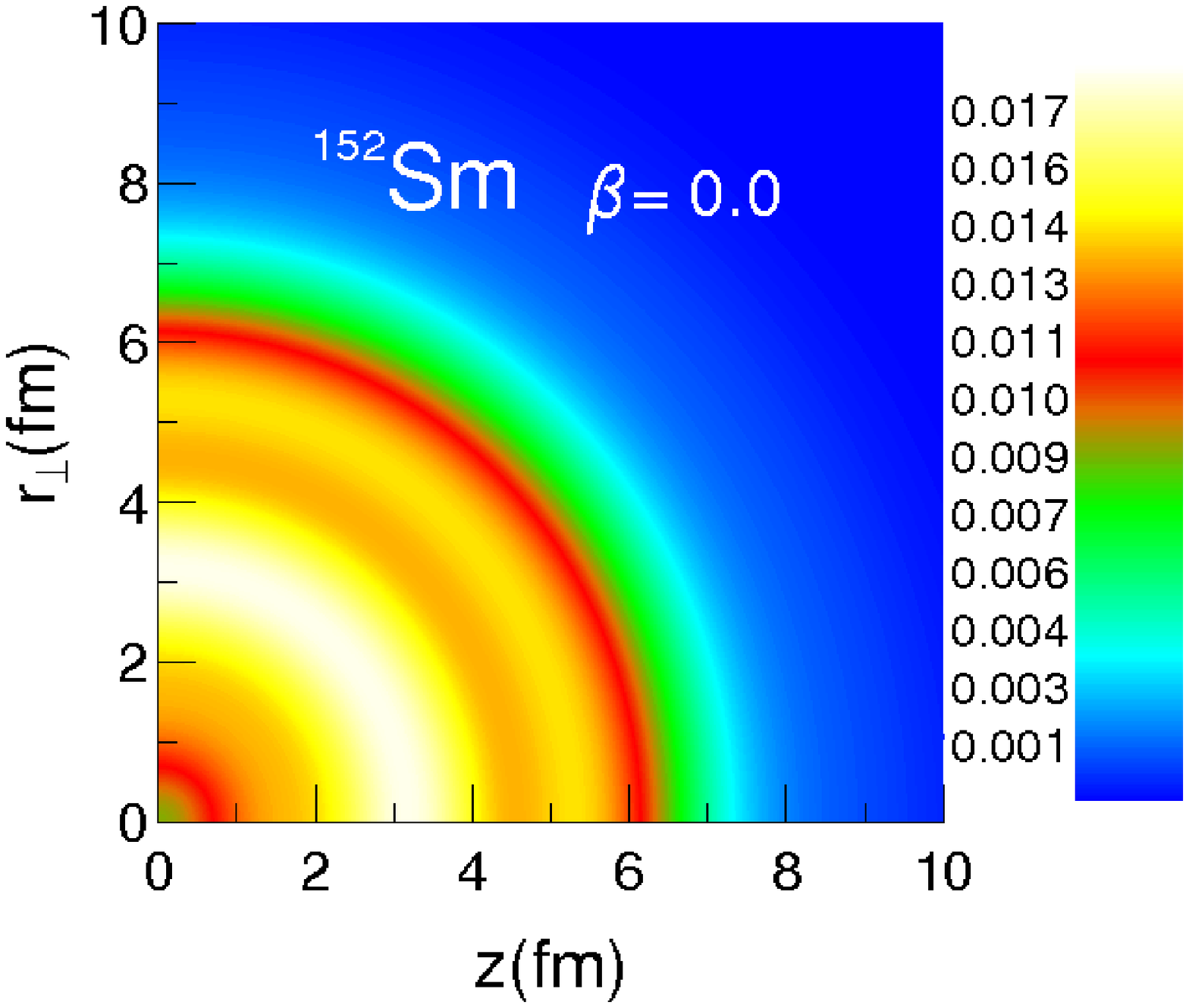}
\includegraphics*[scale=0.45,angle=0]{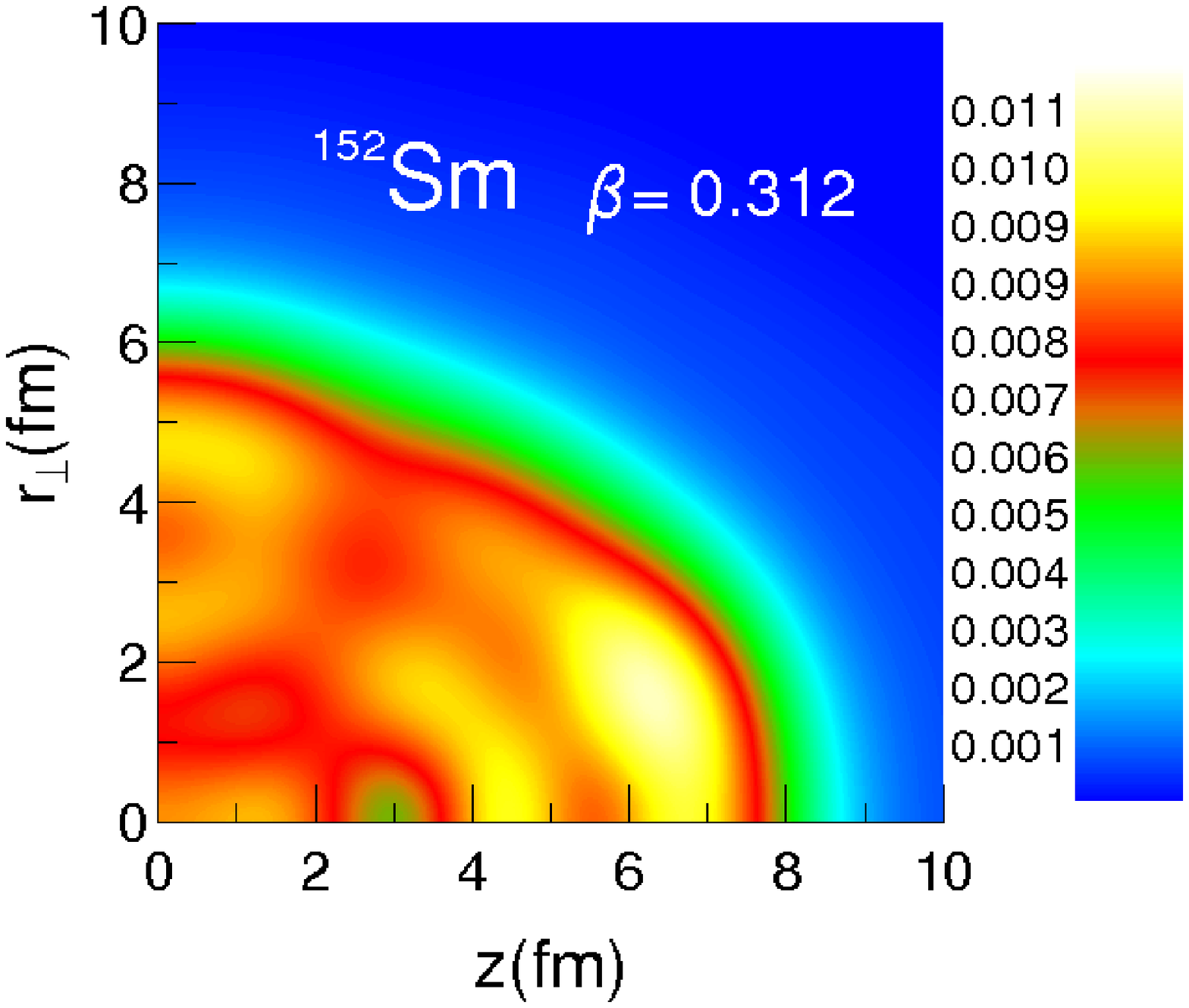}
\end{center}
\caption{ (Color online) The local part of the pairing tensor for $^{152}Sm$.
The upper and lower panels correspond to the spherical
configuration and the deformed ground state, respectively.}
\label{fig4}
\end{figure}

\begin{figure}
\begin{center}
\includegraphics*[scale=0.55,angle=0]{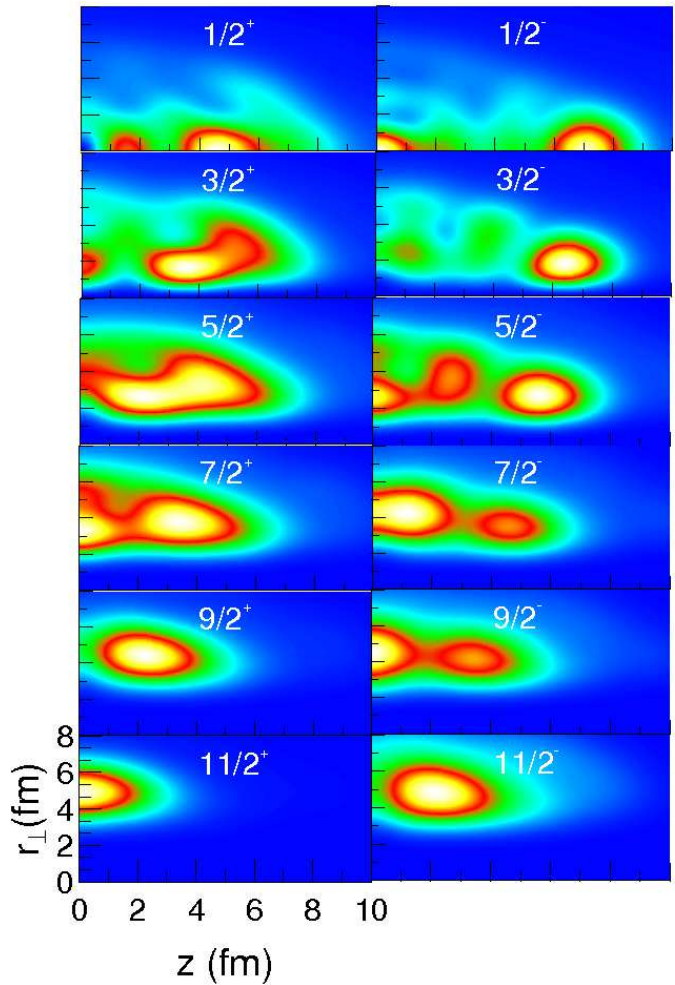}
\caption{ (Color online) The spatial structure of the single-particle blocks
$\Omega^{\pi}$ which have the largest contribution to the pairing tensor shown
in the bottom panel of Fig.\ref{fig4}.}
\label{fig5}
\end{center}
\end{figure}

We shall start by shortly discussing the local part of the
neutron pairing tensor. To illustrate the effect of the
deformation, in Fig.\ref{fig4} is displayed the neutron local part of the pairing tensor for
$^{152}$Sm calculated in the  spherical configuration $\beta =0$ and in the deformed ground
state $\beta=0.312$, where $\beta$ is the usual dimensionless deformation parameter.
The color scaling on the right side of plots indicates the intensity of the local part of the pairing tensor.
In the spherical state the spatial structure of the local part of the pairing tensor can be simply traced back
to the spatial localisation of a few orbitals with energies close to the chemical potential
\cite{sandulescu}. For $^{152}$Sm the most important orbitals are $2f_{7/2}$, $1h_{9/2}$, $3p_{3/2}$
and $2f_{5/2}$. As seen in Fig.\ref{fig4}, in the deformed state the spatial pattern of the pairing
tensor is more complicated. This stems from the fact that it requires quite many single-particle
configurations to explain its detailed structure. The spatial distribution of the configurations
contributing the most to the pairing tensor are shown in Fig.\ref{fig5} ; the plots correspond
to the contribution of single-particle states of given $\Omega$ and parity, with a different scaling
for each panel.

\begin{figure}
\begin{center}
\includegraphics*[scale=0.45,angle=0]{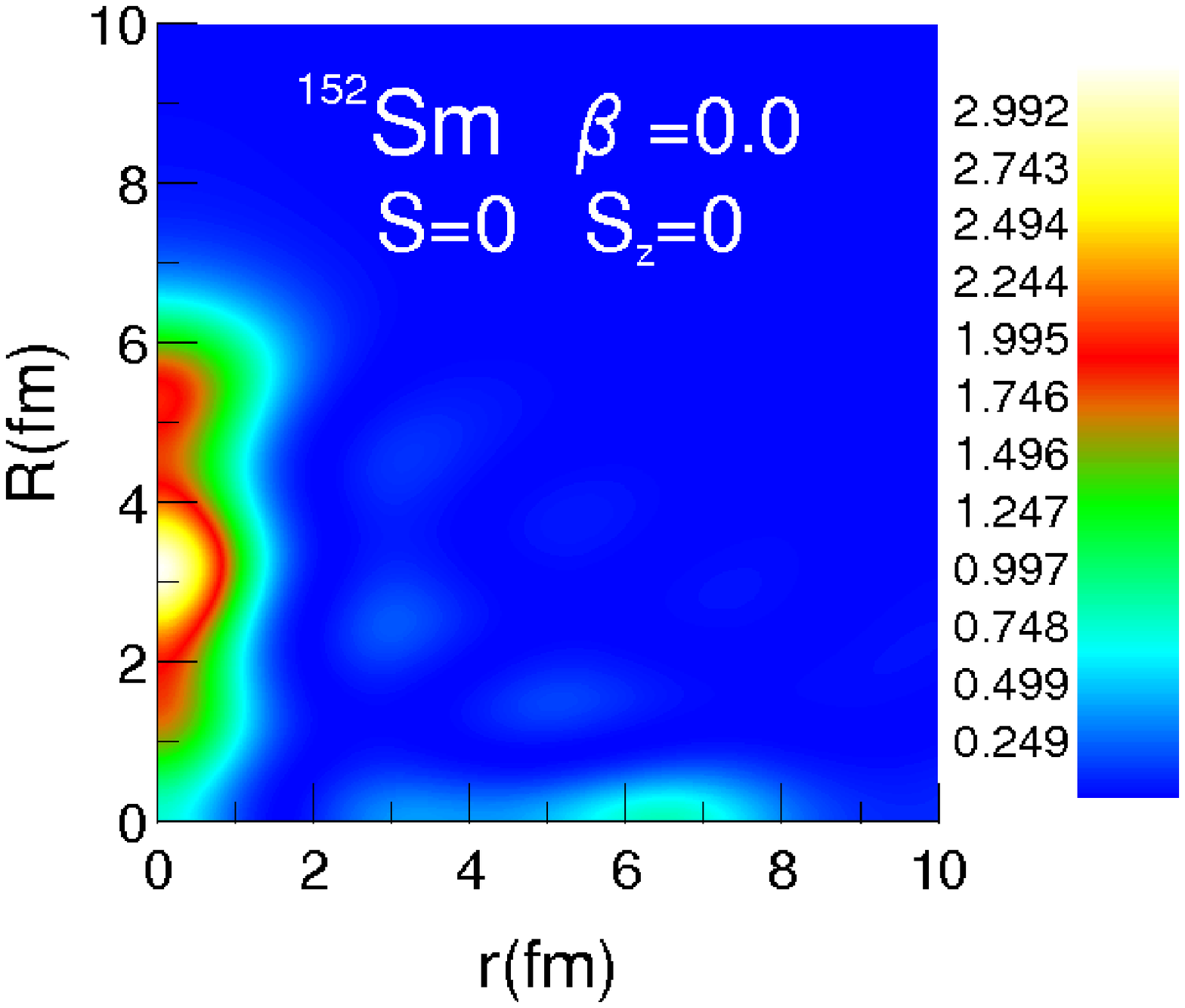}
\includegraphics*[scale=0.45,angle=0]{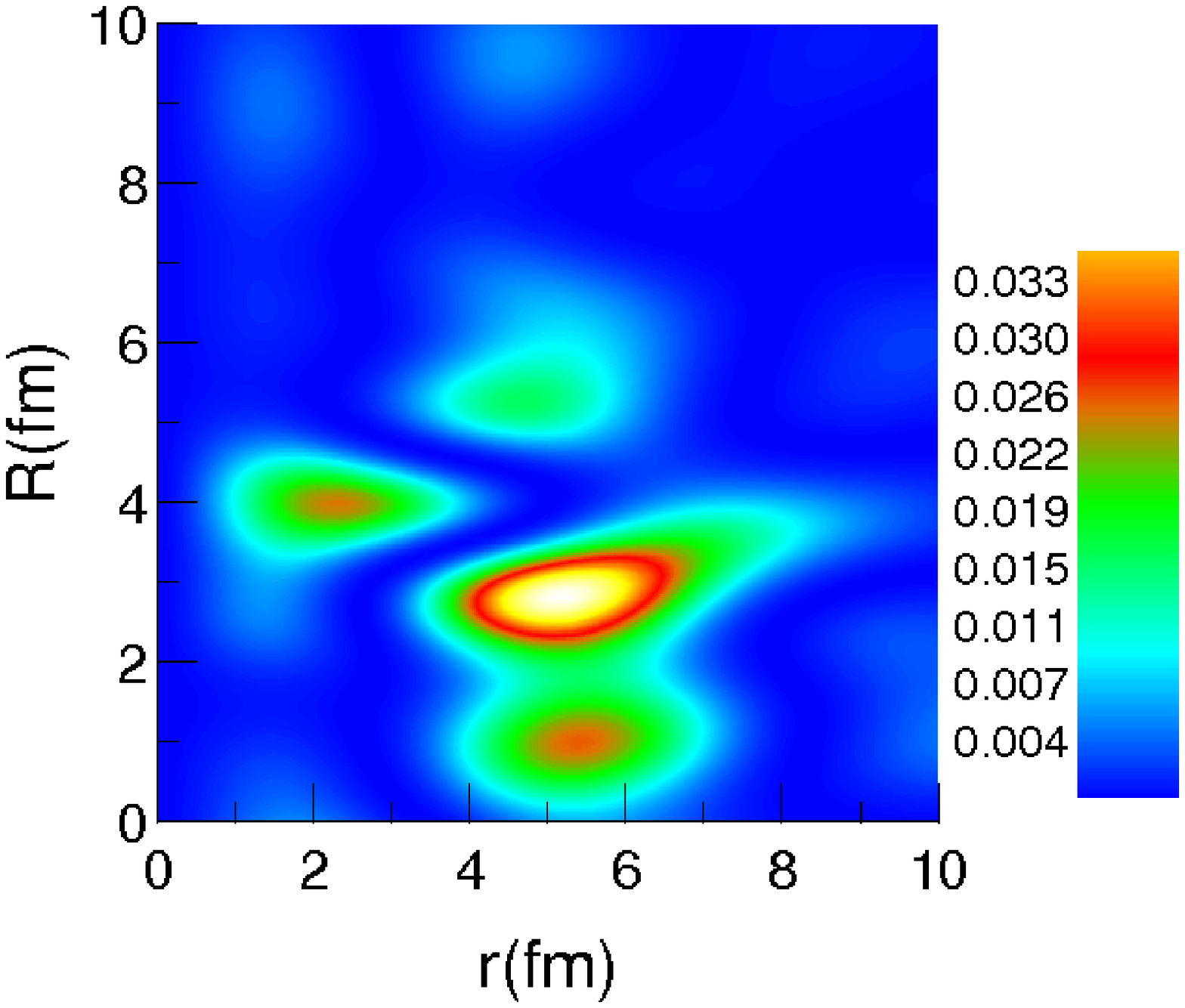}
\end{center}
\caption{ (Color online) Non local $\kappa(\vec{R}, \vec{r})^2$ for the isotope
$^{152}Sm$. The deformation is indicated by $\beta$ and $S$ is the spin.
For the spherical case (upper panel), $\kappa(\vec{R}, \vec{r})$ is averaged over
the angles of $\vec{R}$ and $\vec{r}$. For the deformed case (lower panel), geometrical configuration (a)
of Fig.\ref{fig1} has been adopted.}
\label{fig44}
\end{figure}

\begin{figure}
\begin{center}
\includegraphics*[scale=0.45,angle=0]{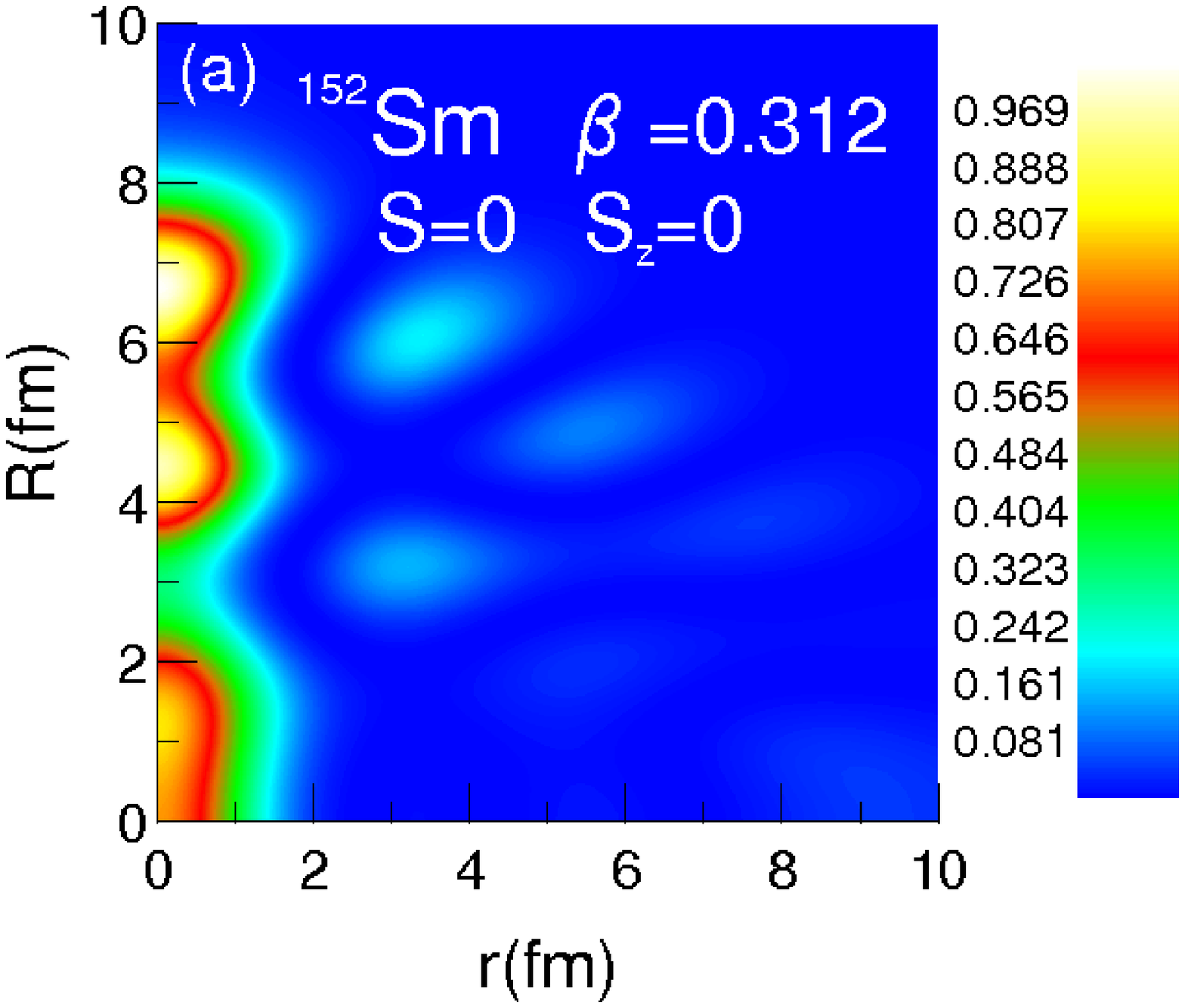}
\includegraphics*[scale=0.45,angle=0]{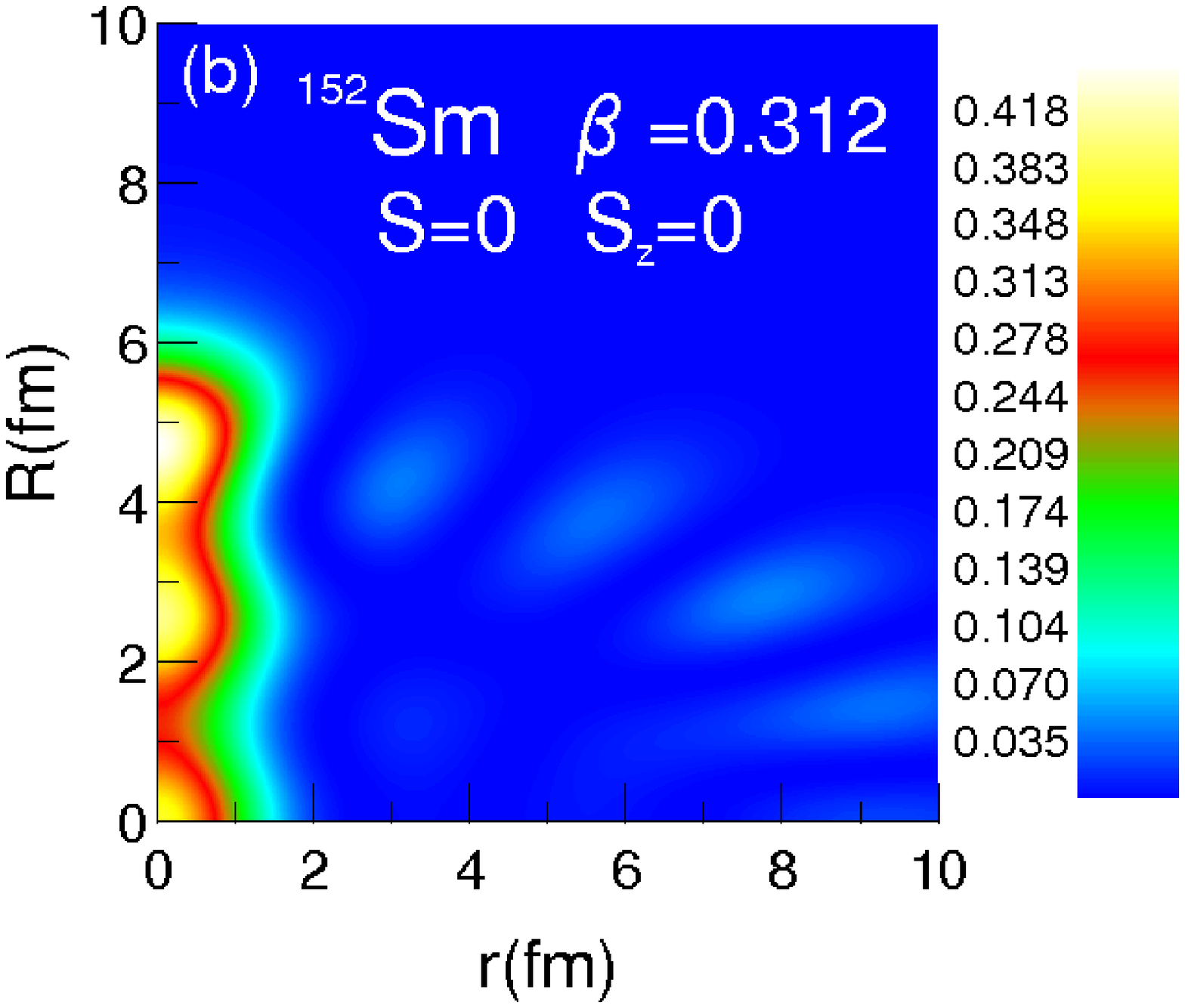}
\includegraphics*[scale=0.45,angle=0]{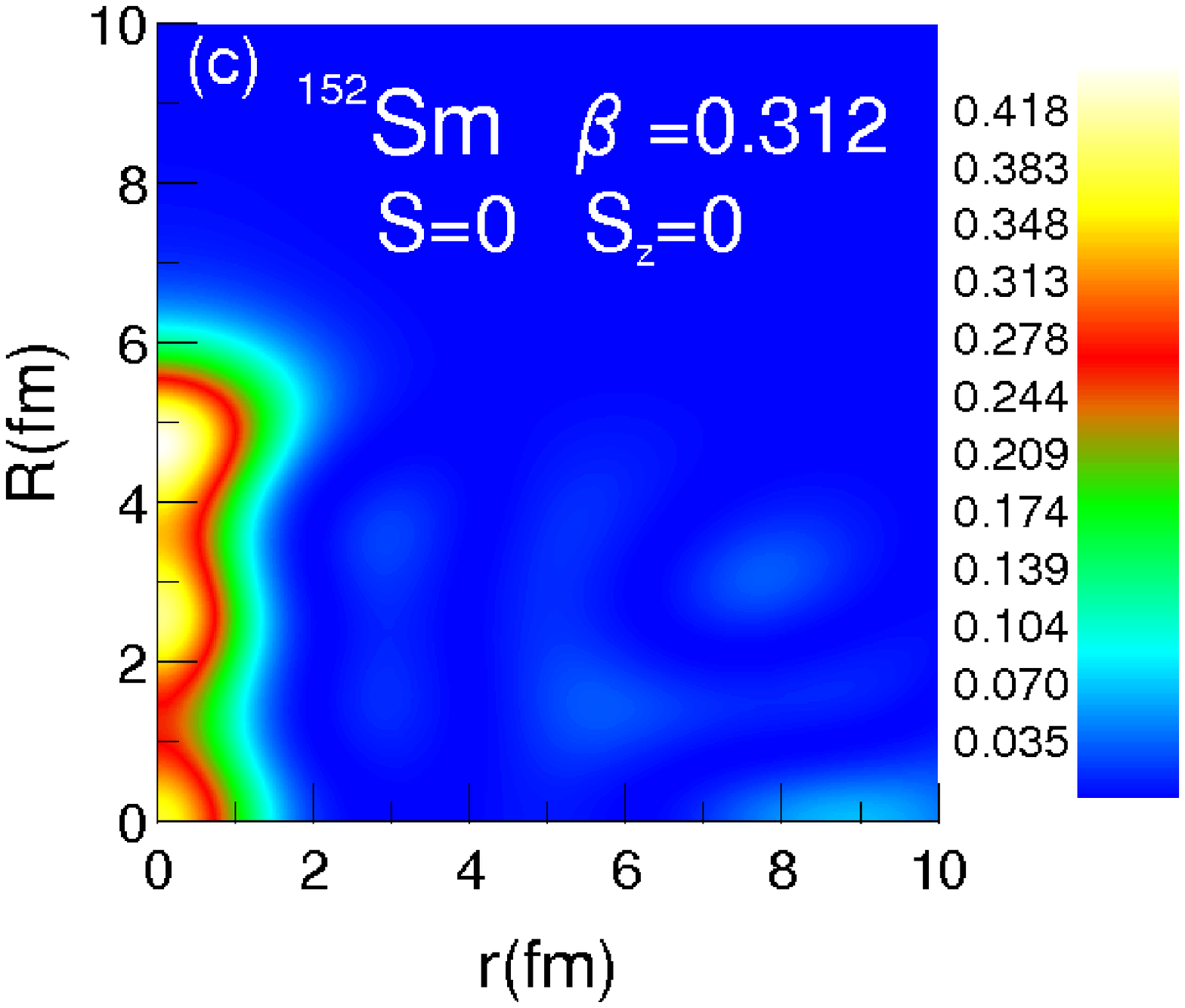}
\end{center}
\caption{ (Color online) Non local $\kappa(\vec{R}, \vec{r})^2$ for the isotope
$^{152}Sm$. The deformation is indicated by $\beta$ and $S$ is the spin.
(a), (b) and (c) indicate the geometrical configurations
shown in Figs \ref{fig1}-\ref{fig3}.}
\label{fig444}
\end{figure}

\begin{figure}
\begin{center}
\includegraphics*[scale=0.45,angle=0]{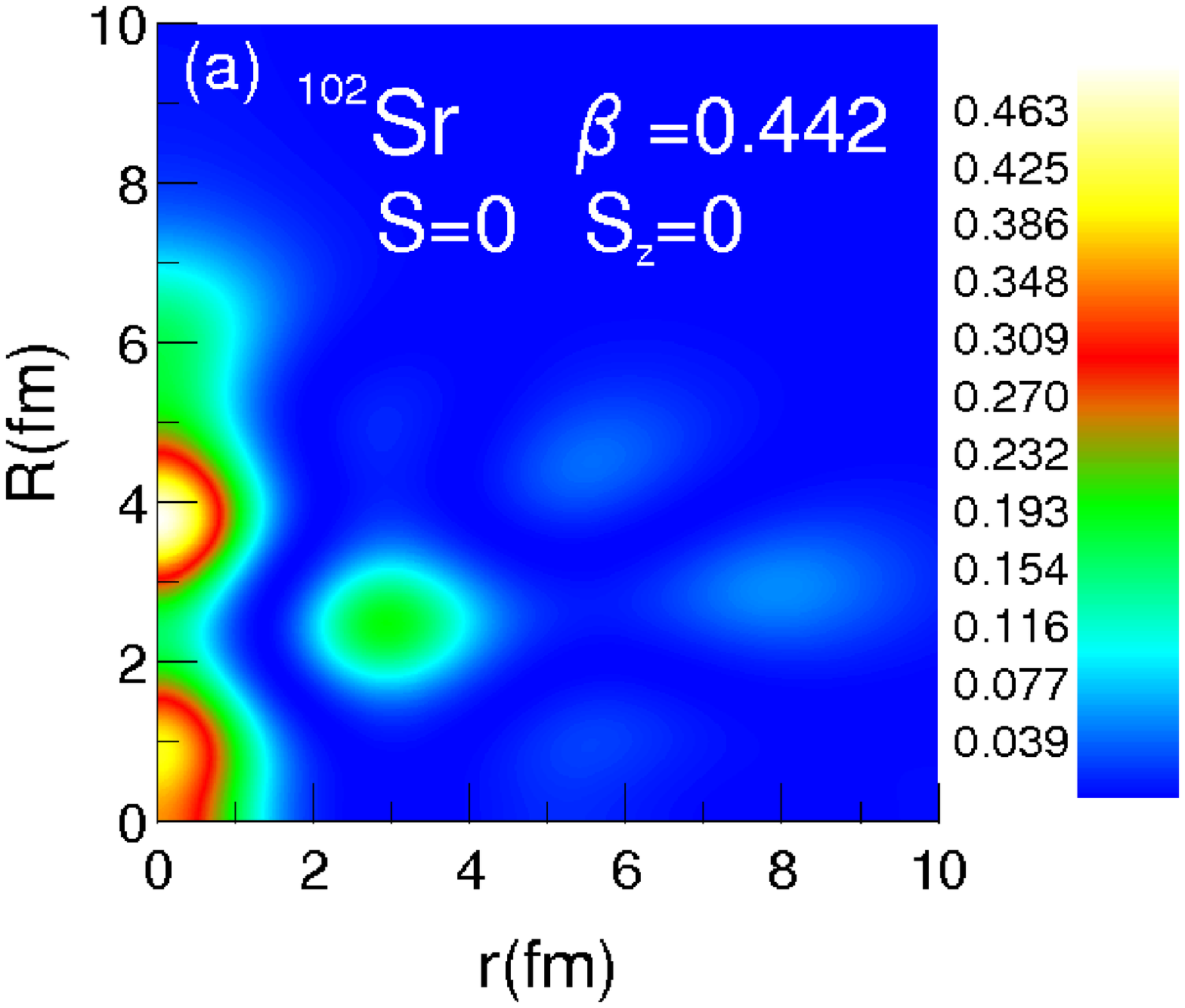}
\includegraphics*[scale=0.45,angle=0]{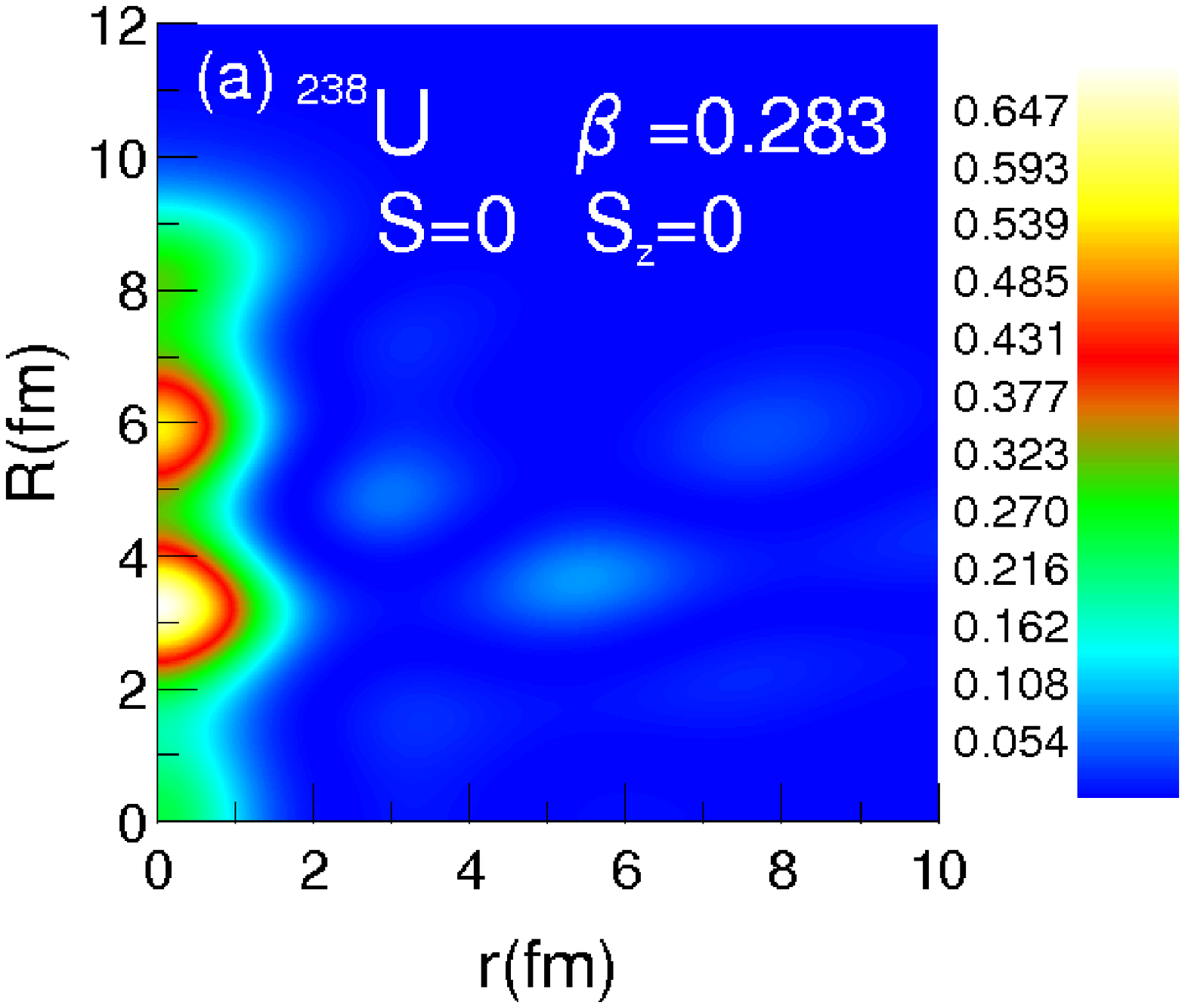}
\includegraphics*[scale=0.45,angle=0]{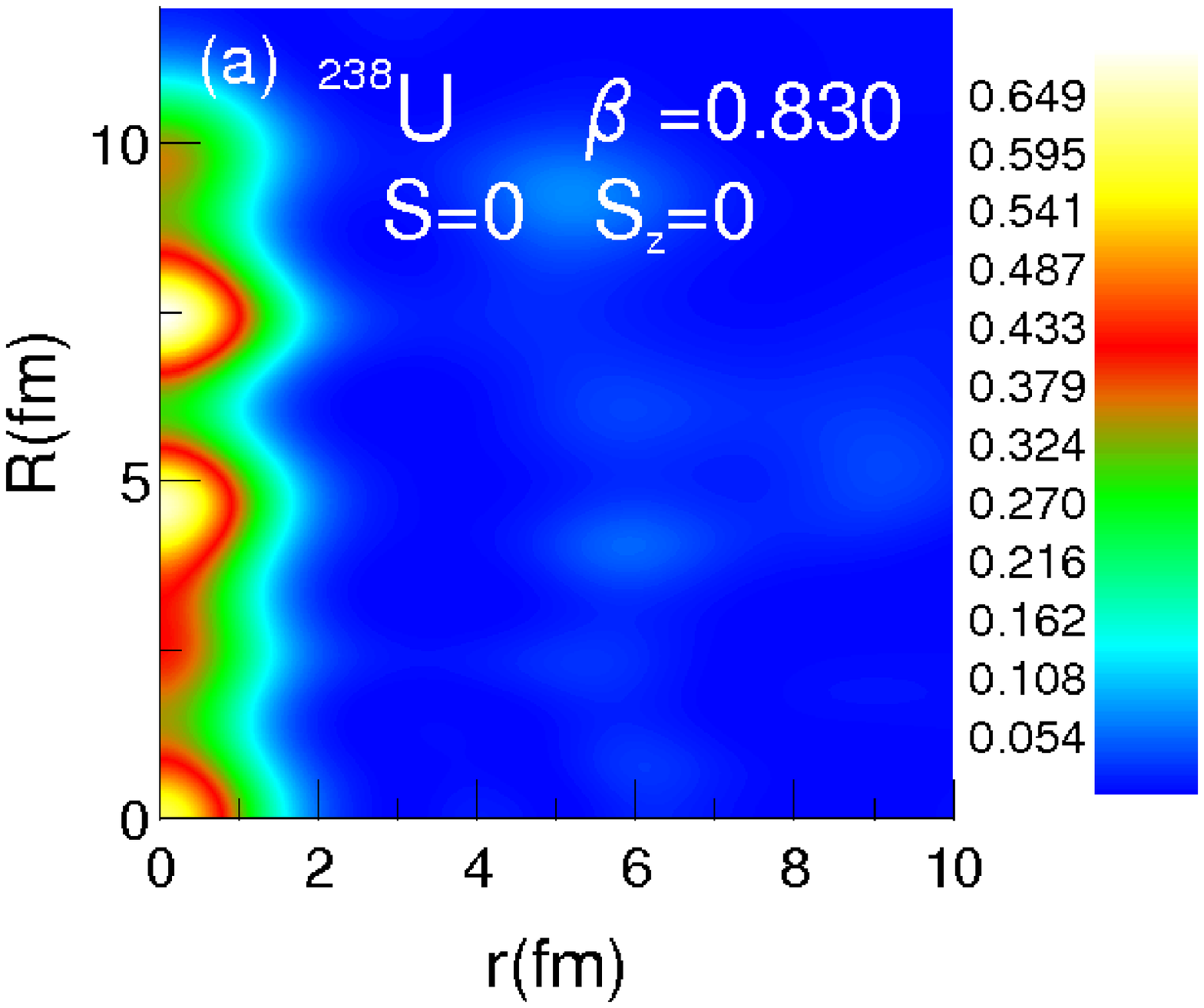}
\end{center}
\caption{ (Color online) Non local $\kappa(\vec{R}, \vec{r})^2$
for isotopes $^{102}Sr$ and $^{238}$U.
For the latter are shown two cases corresponding to the ground
state (middle panel) and to the fission isomer
(bottom panel). Calculations have been made assuming configuration
(a) of Fig.\ref{fig1}.
}
\label{fig66}
\end{figure}

We shall focus now on the spatial structure of the non-local neutron pairing tensor.
In Figs.\ref{fig44}-\ref{fig66} are shown some typical results of
$|\kappa \left( \vec{R}, \vec{r} \right)|^2$ in the three geometrical configurations
(a), (b) and (c) described in Fig.\ref{fig1}-\ref{fig3} and for $^{152}$Sm, $^{102}$Sr and $^{238}$U.
At the spherical deformation, the three geometrical configurations are equivalent.
For $^{102}$Sr that manifests coexistence feature, $|\kappa \left( \vec{R}, \vec{r} \right)|^2$ is
shown only for the prolate minimum. For $^{238}$U, the ground state as well as the isomeric state
are displayed. The color scaling on the right side of plots indicates the intensity of the pairing
tensor squared multiplied by a factor $10^{4}$. First of all we notice  that the pairing tensor for $S=1$
(Fig.\ref{fig44}, bottom panel) is much weaker than for $S=0$ (Fig.\ref{fig444}, top panel)
by a factor $\sim 20$. This is a general feature in open shell nuclei (the pairing channel S=1
is significant in halo nuclei such as $^{11}$Li). Therefore, in what follows we shall discuss
only the channel $S=0$.

Figs.\ref{fig44}-\ref{fig66} show that with deformation the pairing tensor is essentially
confined along the direction of the c.o.mall coordinate. As in spherical
nuclei, the pairing tensor can be preferentially concentrated either in the
surface or in the bulk, depending on the underlying shell structure.
The most interesting fact seen in Figs.~\ref{fig44}-\ref{fig66} is the small spreading of
pairing tensor in the relative coordinate. This is a feature we have already
observed in spherical nuclei. In Ref.\cite{ref1}, it is discussed that
the predilection for small spreading in the relative coordinate is caused by
parity mixing. We will go back to this point in section \ref{3b}.

\noindent

Quantitatively the spreading of the pairing tensor in the relative coordinate can be
measured by the local coherence length (CL) defined in Ref.\cite{blasio}.
In the present study of deformed nuclei, as particular angular dependences
are assumed according to the three geometrical configurations (a), (b) and (c), 
the following formula has been used:
\begin{equation}
\dspt \xi(R) = \sqrt{ \frac
{\int r^4 |\kappa(R,r)|^2 dr}
{\int r^2 |\kappa(R,r)|^2 dr} }
\label{eq20}
\end{equation}
The pairing tensor $\kappa(R,r)$ corresponds to a given total
spin S$=0$ and a given geometrical configuration. 
For spherical nuclear configurations, the expression adopted for the CL is the standard one 
defined in Ref.\cite{ref1} where averages are taken over both the angles of $\vec{R}$ and $\vec{r}$.

\begin{figure}
\begin{center}
\includegraphics*[scale=0.45,angle=0]{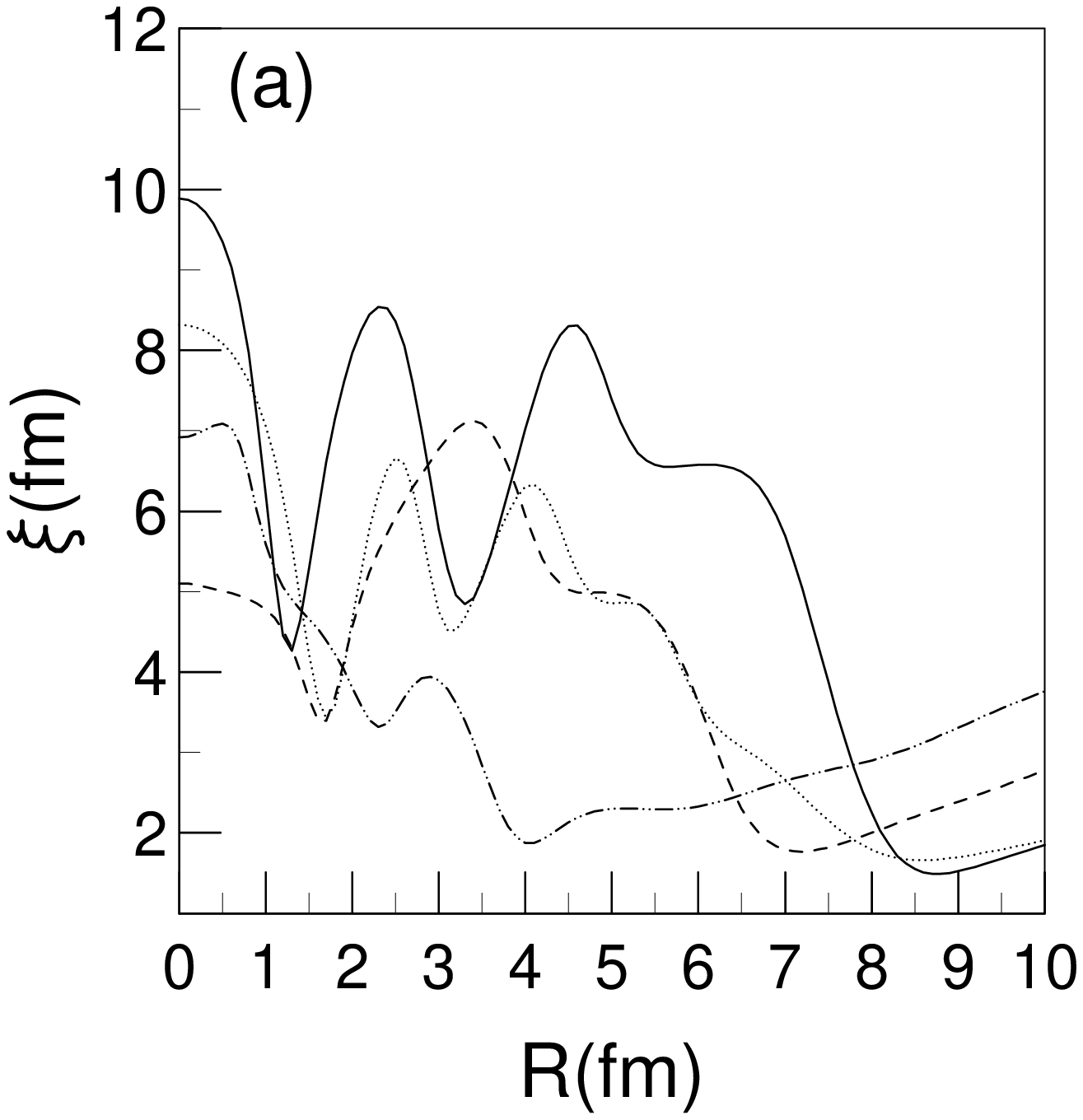}
\includegraphics*[scale=0.45,angle=0]{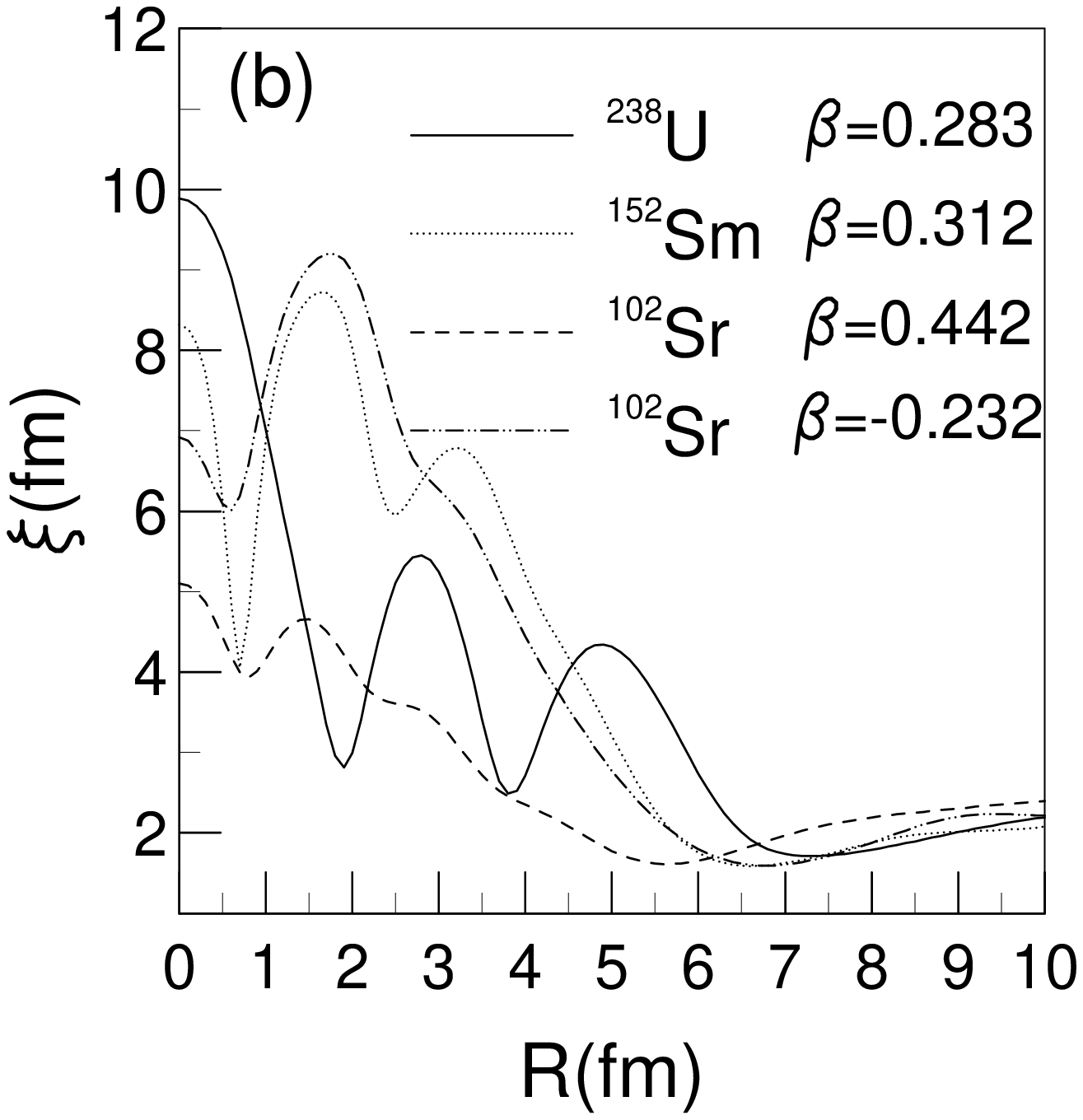}
\includegraphics*[scale=0.45,angle=0]{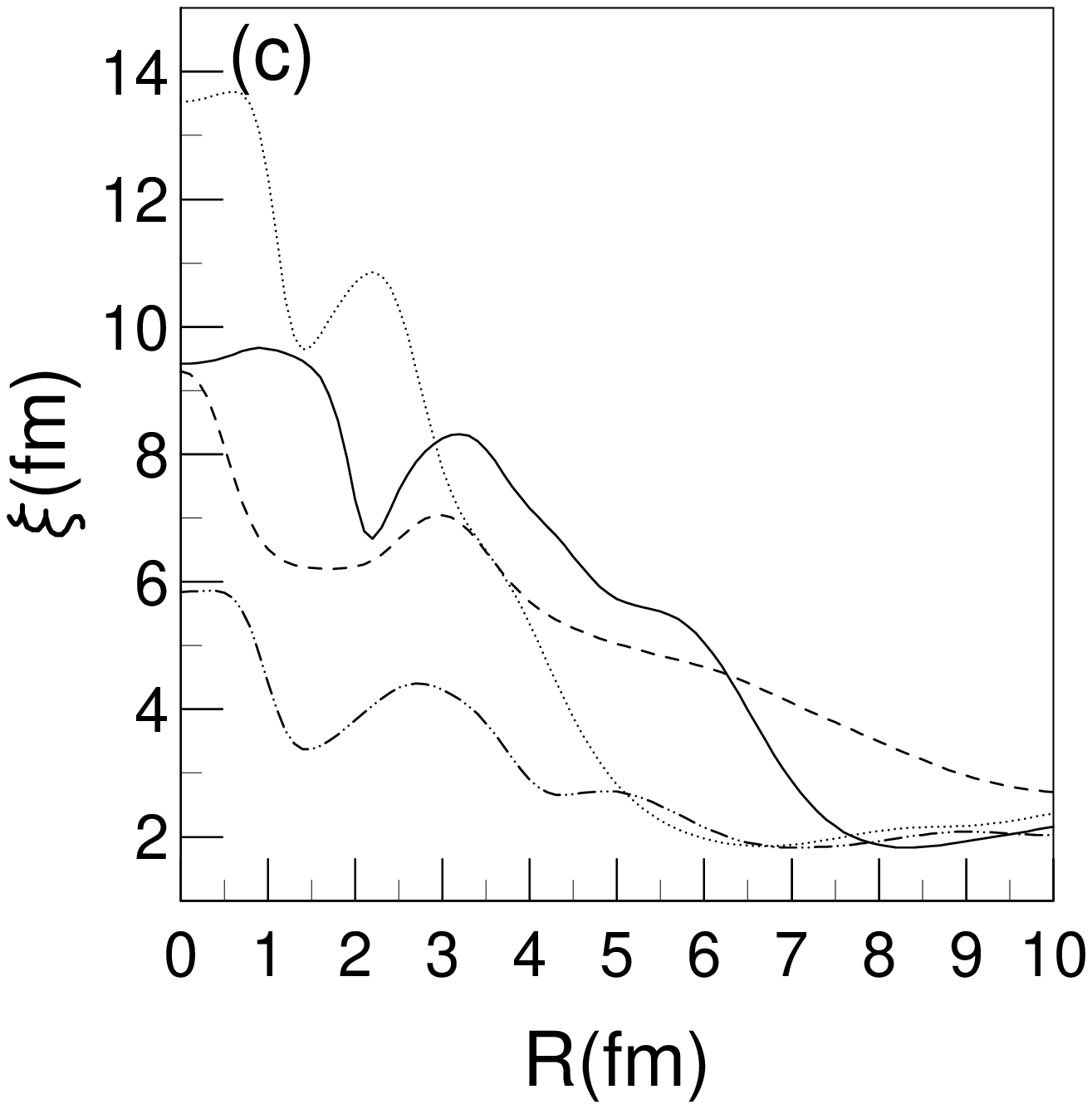}
\caption{ Coherence length for various isotopes. $\beta$ denotes
the deformation while (a),(b),(c) are the geometrical configurations
shown in Figs. \ref{fig1}-\ref{fig3}.}
\label{fig8}
\end{center}
\end{figure}

In Fig.\ref{fig8}, we present the neutron CL calculated for various
deformed nuclei and configurations (a), (b) and (c) described in Fig.\ref{fig1}-\ref{fig3}.
We notice that inside the nucleus the CL has large values, up to about 10-14 fm. This order of
magnitude was already found in spherical nuclei. However, the CL displays
much stronger oscillations compared to spherical nuclei, especially for
the geometrical configurations (a) and (b). This behaviour can be attributed to the large number
of different orbitals implied in pairing properties for deformed nuclei.
An interesting feature seen in Fig.\ref{fig8} is the pronounced minimum of about 2 fm far
out in the surface which appears for all isotopes and all geometrical configurations.
The minimum found here has a similar magnitude as in spherical nuclei. A small coherence length
of $\sim$ 2 fm in the surface of nuclei we have also found for the protons. In the proton case,
the Coulomb force has not been taken into account in the pairing interaction but it is not expected
to change the CL strongly.

\begin{figure}
\begin{center}
\includegraphics*[scale=0.45,angle=0]{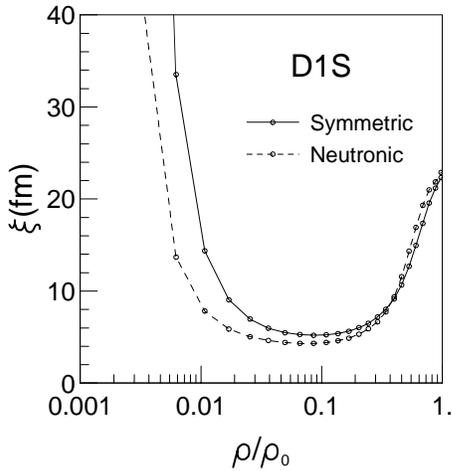}
\end{center}
\caption{ Coherence length in symmetric and neutronic matter according to the density normalized to the saturation
density and calculated with the D1S Gogny force.}
\label{fig100}
\end{figure}

\subsection{Discussion of the coherence length} \label{3b}

Compared to the smallest values of the CL in nuclear matter, of
about 4-5 fm (see Fig.\ref{fig100} for symmetric and neutronic matter),
the minimal values ($\sim$ 2 fm) of the CL in nuclei are astonishingly small.
The question which then arises is what causes such small values
of the CL in the surface of nuclei.
Since, as we have just mentioned, the general behaviour of the CL is
similar in spherical and deformed
nuclei, in what follows we shall focus the discussion on spherical
nuclei. As a benchmark case, we will consider the isotope $^{120}$Sn.
In this case, the CL will be calculated as in Ref.\cite{ref1},
\begin{equation}
\dspt \xi(\vec{R}) = \sqrt{ \frac
{\int r^2 |\kappa(\vec{R},\vec{r})|^2 d^3r}
{\int |\kappa(\vec{R},\vec{r})|^2 d^3r} }
\label{eq20b}
\end{equation}
where averages are taken over both the angles of $\vec{R}$ and $\vec{r}$.

A possibility explaining the small CL in the surface of finite nuclei could be, as, e.g., suggested
in \cite{ref1}, that pairing correlations are particularly strong there.
Indeed, in the surface the neutron Cooper pairs have approximatively the same size as the deuteron, a bound pair.
This is a situation similar to the strong coupling regime of pairing correlations. However, it is generally 
believed that nuclei are with respect to pairing in the weak coupling limit \cite{bm}.
In what follows, we shall examine whether there exists a correspondence
between the magnitude of the CL and an enhancement of pairing correlations in the nuclear surface.
Even though a local view can only give an incomplete picture because of fluctuations, a quantity which can
be used to explore the spatial distribution of pairing correlations is the local pairing energy

\begin{equation}
E_c(R) =  -\int d^3r \Delta(\vec{R},\vec{r}) \kappa (\vec{R},\vec{r}) ,
\label{ec}
\end{equation}
\noindent
where $\Delta(\vec{R},\vec{r})$ is the nonlocal pairing field. In practice in
Eq.(\ref{ec}), we use the angle-averaged quantities.
\begin{figure}
\begin{center}
\vspace{-0.75cm}
\includegraphics*[scale=0.45,angle=0]{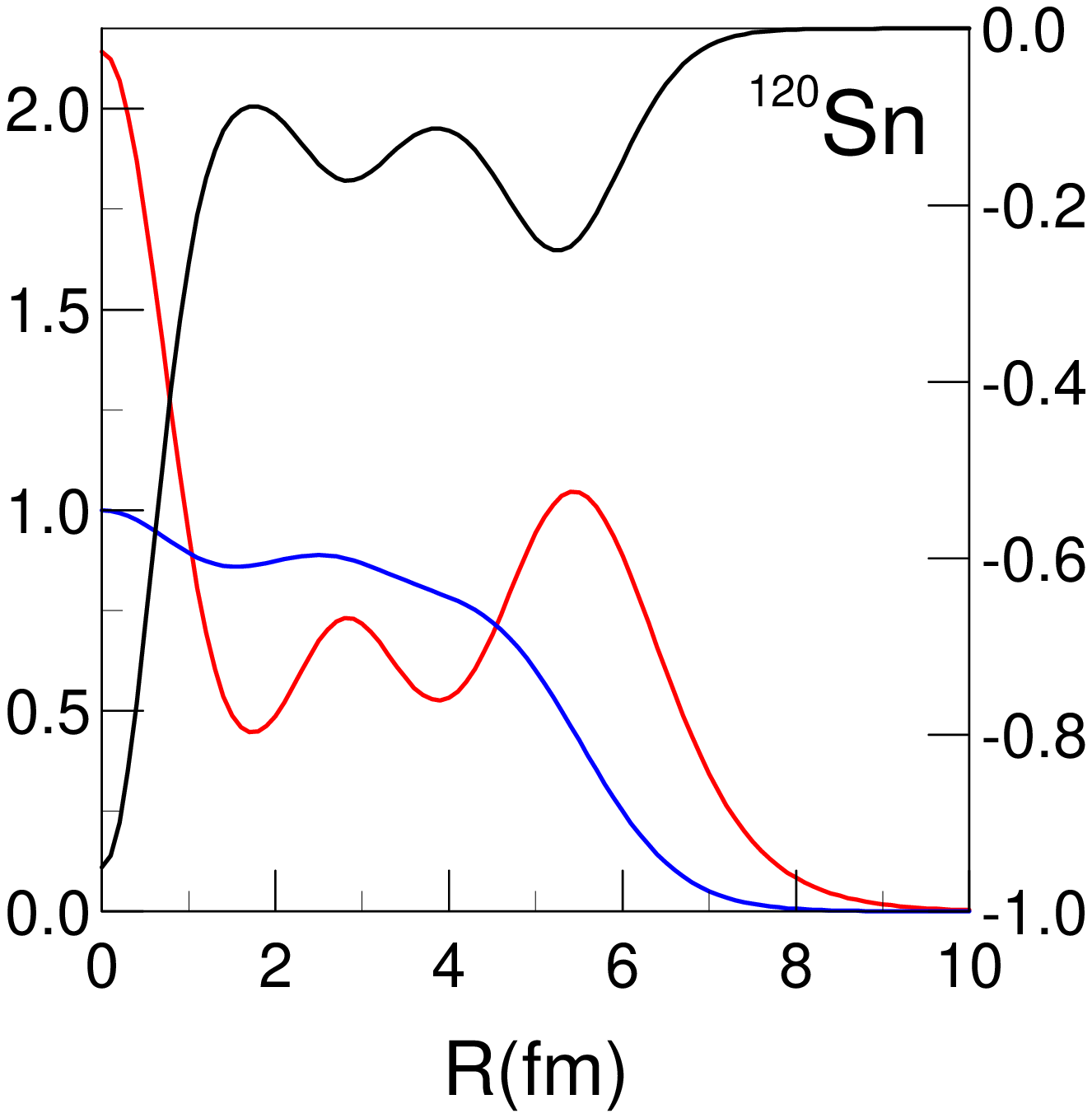}
\vspace{-0.75cm}
\includegraphics*[scale=0.45,angle=0]{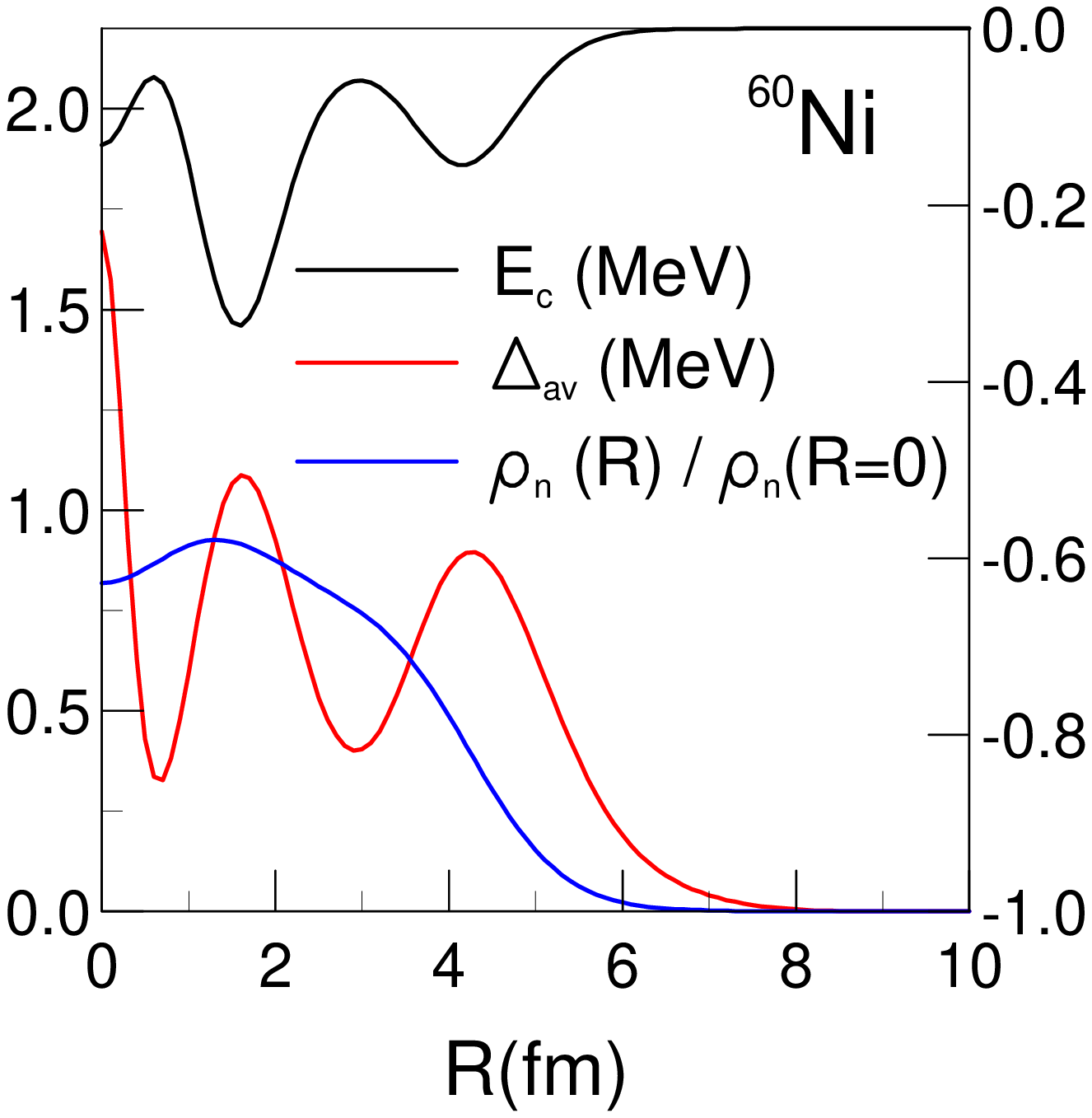}
\vspace{-0.75cm}
\includegraphics*[scale=0.45,angle=0]{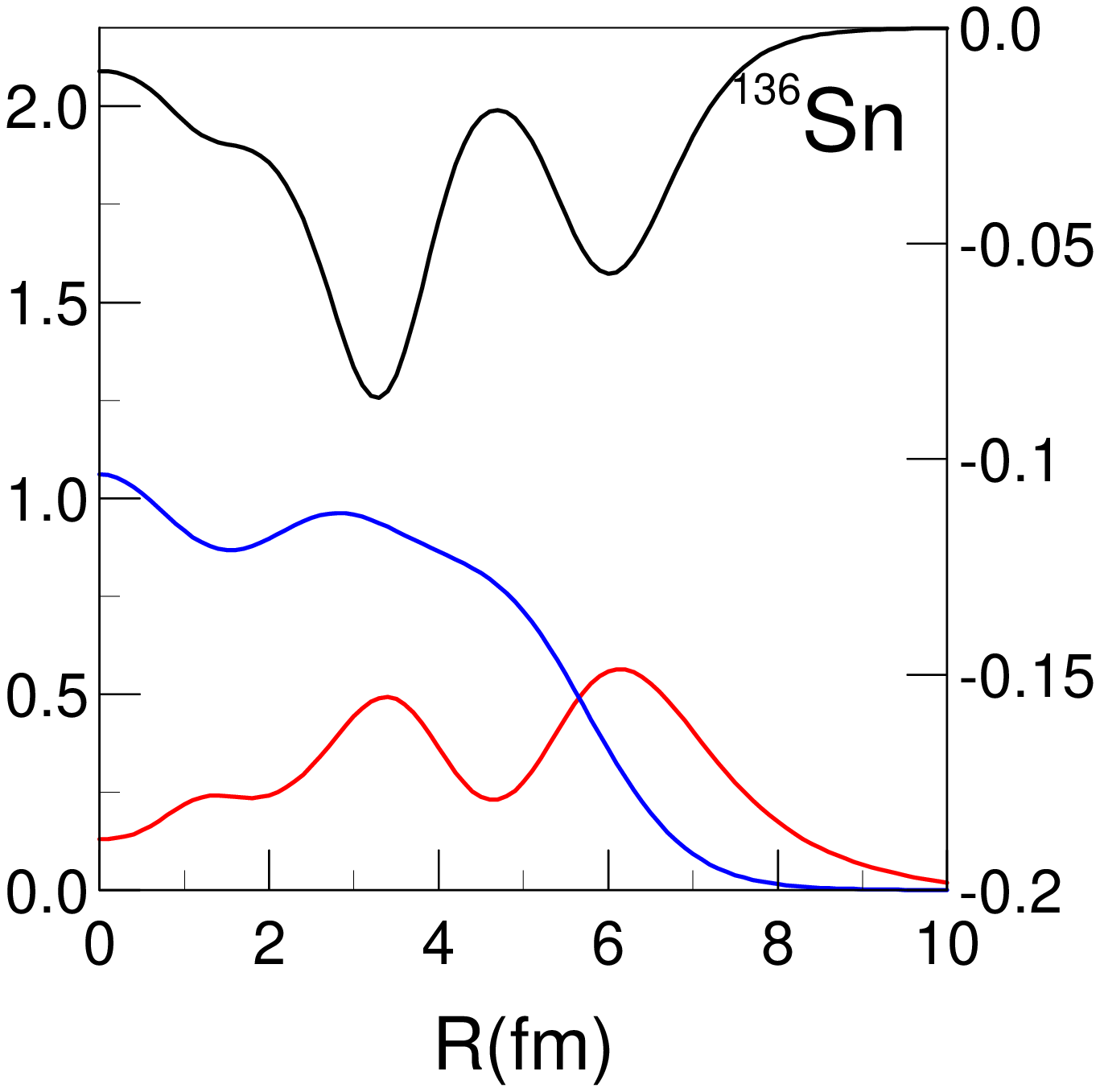}
\vspace{-0.75cm}
\includegraphics*[scale=0.45,angle=0]{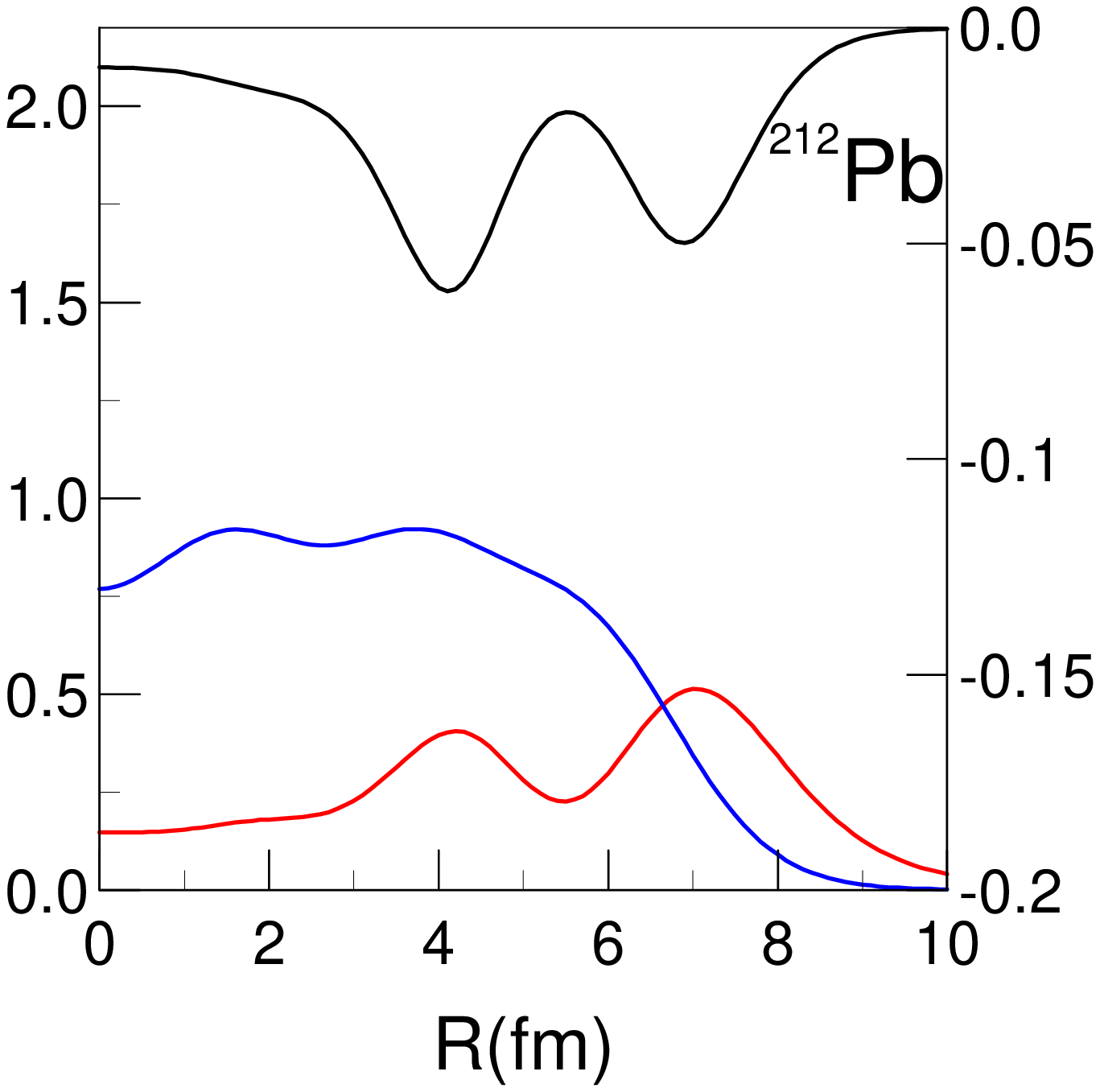}
\end{center}
\caption{ (Color online) Pairing correlation energies (right scale in MeV) and the average pairing
fields (left scale in MeV) in $^{120}$Sn, $^{60}$Ni, $^{136}$Sn and $^{212}$Pb.
By dashed-dotted line is shown the neutron density relative to its value
in the center of the nucleus (left scale).}
\label{fig9}
\end{figure}

The localisation properties of $E_{c}(R)$ can be seen in Fig.\ref{fig9} (black line) where we show the
results for several spherical nuclei. We notice that in the surface region where the minimum of
the CL is located there is a local maximum of $|E_{c}(R)|$ present for all nuclei considered.
The largest value of $E_{c}(R)$ (in absolute value) is not necessarily located in the surface region
and the oscillations of the inner part of the distributions seem mostly due to shell fluctuations.

In order to better exhibit a surface enhancement of pairing correlations, we have to consider a normalized
pairing energy, otherwise the strong fall off of the density will mask to a great deal the local increase of
pairing. One could divide $E_c(R)$ by the local density, as done in \cite{tischler}. However, here we prefer
to divide by the local pairing density $\kappa(R) = \kappa(R,0)$, leading to the following definition of
an average local pairing field
\begin{equation}  \label{22}
\dspt \Delta_{av}(R) = \frac{1}{\kappa(R)}\int d^3 r \Delta({\vec R},{\vec r})
\kappa ({\vec R},{\vec r})
\end{equation}
\noindent
In practice in Eq.(\ref{22}) we again use the angle-averaged quantities.
We remark that with a zero range pairing force the above definition of the average
pairing field gives the local pairing field.
The localisation properties of $\Delta_{av}(R)$ can be seen in Fig.\ref{fig9} (red line).
We notice a qualitatively similar behavior as for $E_c(R)$.
However, due to the normalization, the average pairing field has significant values
out in the surface region. A closer inspection of Fig.\ref{fig9} shows that the averaged
pairing field reaches by 20$\%$ further out into
the surface of the nucleus compared to the particle density (blue line). This can be quantified by the
corresponding root mean squared values. This push of pairing correlations to the external region is determined
by the localization properties of orbitals from the valence shell, which give the main contribution to
the pairing tensor and pairing field. Since these states are less bound they are more spatially extended
than the majority of states which determine the particle density and the nuclear radius. Moreover,
the increase of the effective mass in the surface also probably plays an important role.
Like the local pairing energy $E_c(R)$, the average pairing field
$\Delta_{av}(R)$ presents a generic local maximum in the surface
region with a local enhancement of pairing correlations (at tenth the matter density in $^{120}$Sn, the average
pairing field still reaches a relatively large value of $\sim 0.5$MeV). On the other side,
this maximum is not necessarily an absolute one and higher pairing field values can appear
in the interior of nuclei, depending on the underlying shell structure.
\begin{figure}
\includegraphics*[scale=0.45,angle=0]{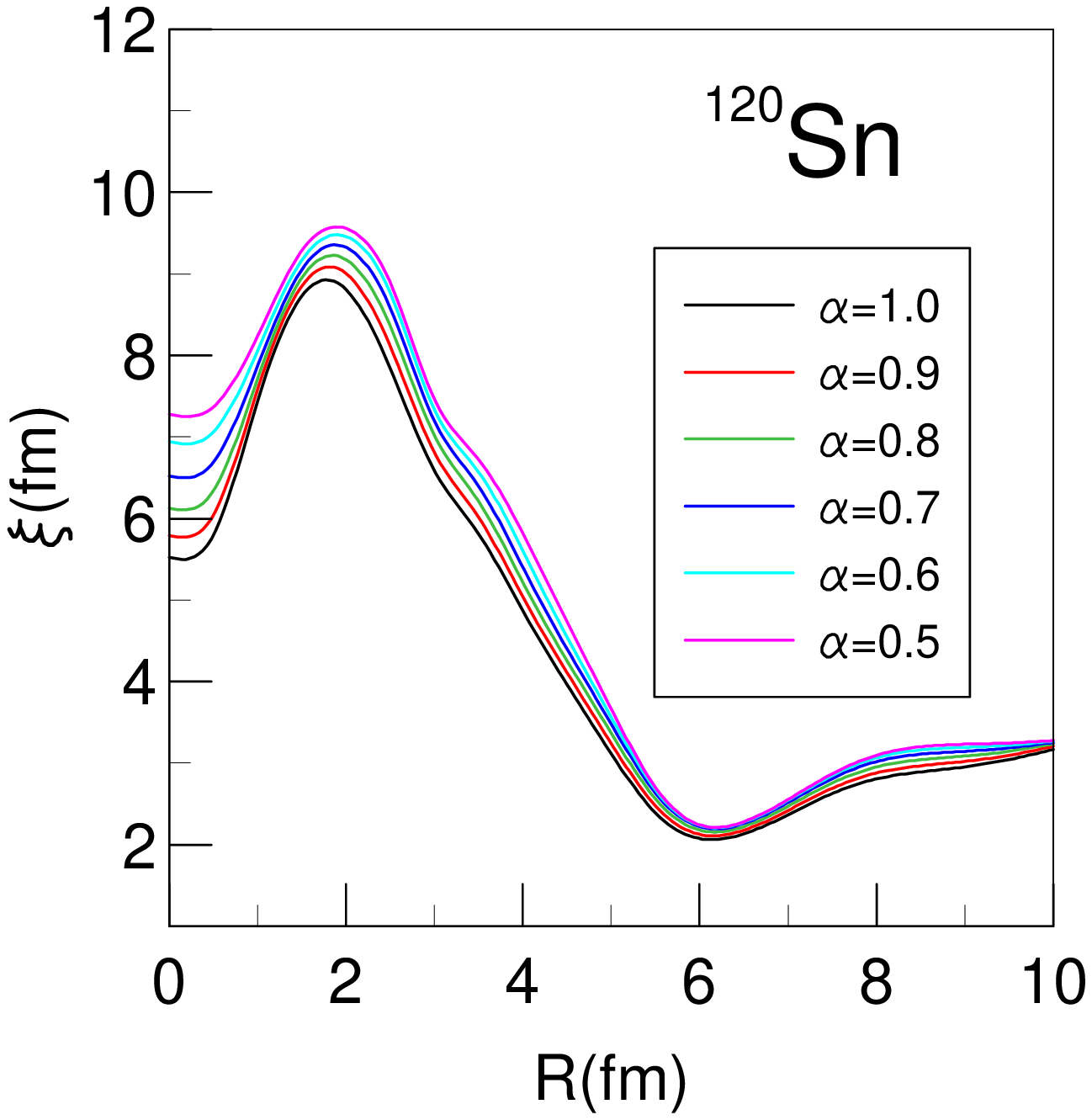}
\includegraphics*[scale=0.45,angle=0]{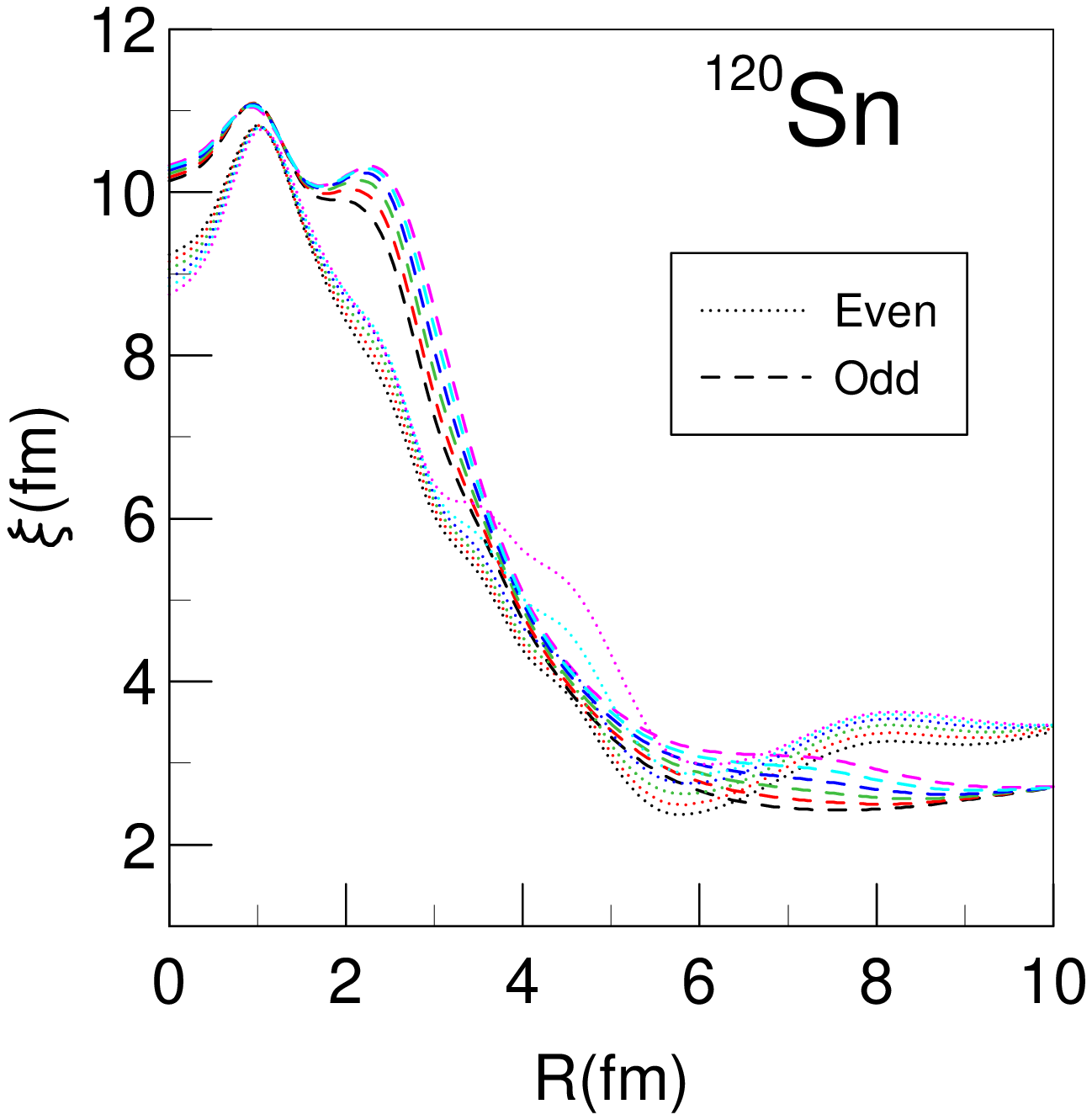}
\caption{ (Color online) Coherence length calculated with total pairing tensor $\kappa$ (top panel),  
even $\kappa_{e}$
and odd $\kappa_{o}$ part of the total pairing tensor (bottom panel) for different intensity of 
pairing strength, in
the case of $^{120}$Sn.}
\label{intensity}
\end{figure}\\

In order to understand if this local enhancement of pairing correlations is
able to explain the minimum value of
$\sim$ 2 fm of the CL in the surface of finite nuclei, we have calculated the
CL under the same conditions as before
but with a variable factor $\alpha$ in front of the (S=0, T=1) pairing
intensity of the D1S Gogny force
(and only there). The result is shown in Fig.\ref{intensity}
for $\alpha$ between 1.0 and 0.5 (top panel) for
$^{120}$Sn. It should be mentioned that for
$\alpha =1.0$, the $^{120}$Sn pairing energy is equal to $\sim 19$MeV
whereas for $\alpha =0.5$ it is $\sim 0.5$MeV which can be considered as a
very weak pairing regime.
In spite of these extreme variations of the pairing field,
the values of the CL are changing overall very little, except for $R \le 1$ fm.
At $R \simeq 6$ fm, the variation
is less than $0.2$ fm. As we will see, this behavior is completely different in
nuclear matter.
From this study, it becomes clear that the CL is practically independent of
the pairing intensity, in
particular in the surface of finite nuclei.

Therefore, we must revisit the interpretation proposed in our preceding 
paper \cite{ref1}, that
the minimal size of $\sim$ 2 fm of Cooper pairs in the nuclear 
surface is a consequence of
particularly strong local pairing correlations.
From the fact that a completely different behavior is obtained in infinite nuclear matter (see below
and Fig.\ref{xinm}), the small size of the CL in the surface of nuclei seems
to be strongly related to the finite size of the nucleus.

At this stage of our analysis, it is important to clarify the role of parity mixing which was put forward in our
preceding work \cite{ref1} on the behaviour of the CL.
In the bottom panel of Fig.\ref{intensity}, is displayed the CL calculated
either with the even part of the pairing
tensor $\kappa_{e}$ or with the odd one $\kappa_{o}$, for the same values of
$\alpha$ as before. One sees that,
in both cases, the value of the CL does not depend much on the intensity of
the pairing. This conclusion holds here for all the values of R.
Comparing the curves in the two panels of Fig.\ref{intensity}, one sees that
the even/odd CL's have sensibly larger values in
the center of the nucleus (around $\sim 10$ fm) than the CL calculated with
the full $\kappa$ (6-8 fm), almost
independently of the value of $\alpha$. In the surface region they
are practically of the same magnitude (2-3 fm). These results indicate that the parity 
mixing discussed in Ref. \cite{ref1}
influences the CL essentially for small values of R. Therefore, parity mixing cannot be the main 
reason for the small value of the CL in the surface region.

\begin{figure}
\includegraphics*[scale=0.55,angle=0]{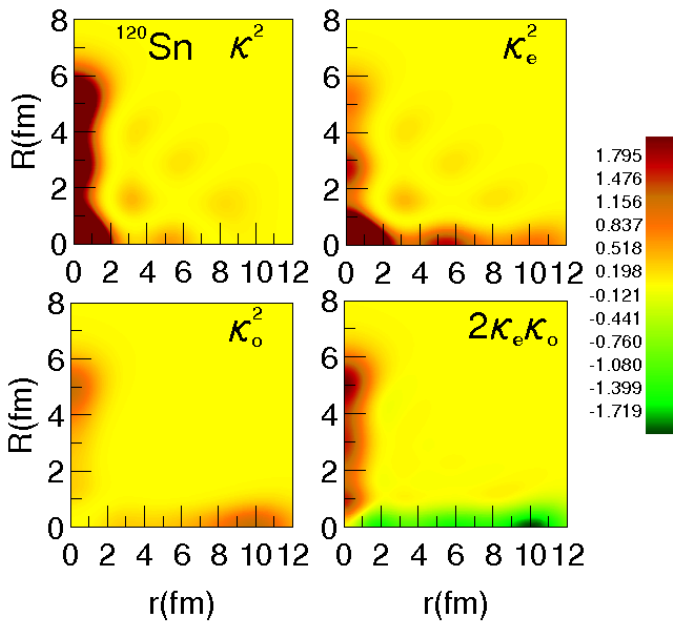}
\caption{(Color online) Non-local part of the pairing tensor $\kappa^2$, even $\kappa_{e}^{2}$
and odd $\kappa_{o}^2$ part of the non-local part of the pairing tensor and the interference term
$2\kappa_{e}\kappa_{o}$ for $^{120}$Sn. }
\label{new3}
\end{figure}

The trends observed in Fig.\ref{intensity} can be traced back to the variations of
$\kappa^2$ as well as $\kappa_{e}^{2}$, $\kappa_{o}^{2}$ and the
interference term $2\kappa_{e} \kappa_{o}$ plotted in Fig.\ref{new3}.
One sees that the interference term is large only along the axes $r=0$ and $R=0$.
However, in calculating the CL, $|\kappa|^2$ is multiplied by a factor $r^4$ in the numerator of Eq.(\ref{eq20b}).
Hence, the large values of this interference term near $r=0$ axis will not come into play significantly.
Therefore, as observed previously,
parity mixing will be significant essentially for $R \lesssim 1$ fm.

This observation is confirmed by looking at the quantity
\begin{equation}
\dspt X(R,r)= \frac{r^{4} |\kappa \left( R, r \right)|^{2}}{N(R)}
\label{enew2}
\end{equation}
where $N(R)=\int_{0}^{\infty} dr~r^2 \kappa(R,r)^2$. This quantity, once integrated over $r$ yields the
square of the CL, Eq.(\ref{eq20b}) namely, $\xi^2(R)= \int_{0}^{\infty} X(R,r) dr$.

\begin{figure}
\vspace{-1cm}
\includegraphics*[scale=0.45,angle=0]{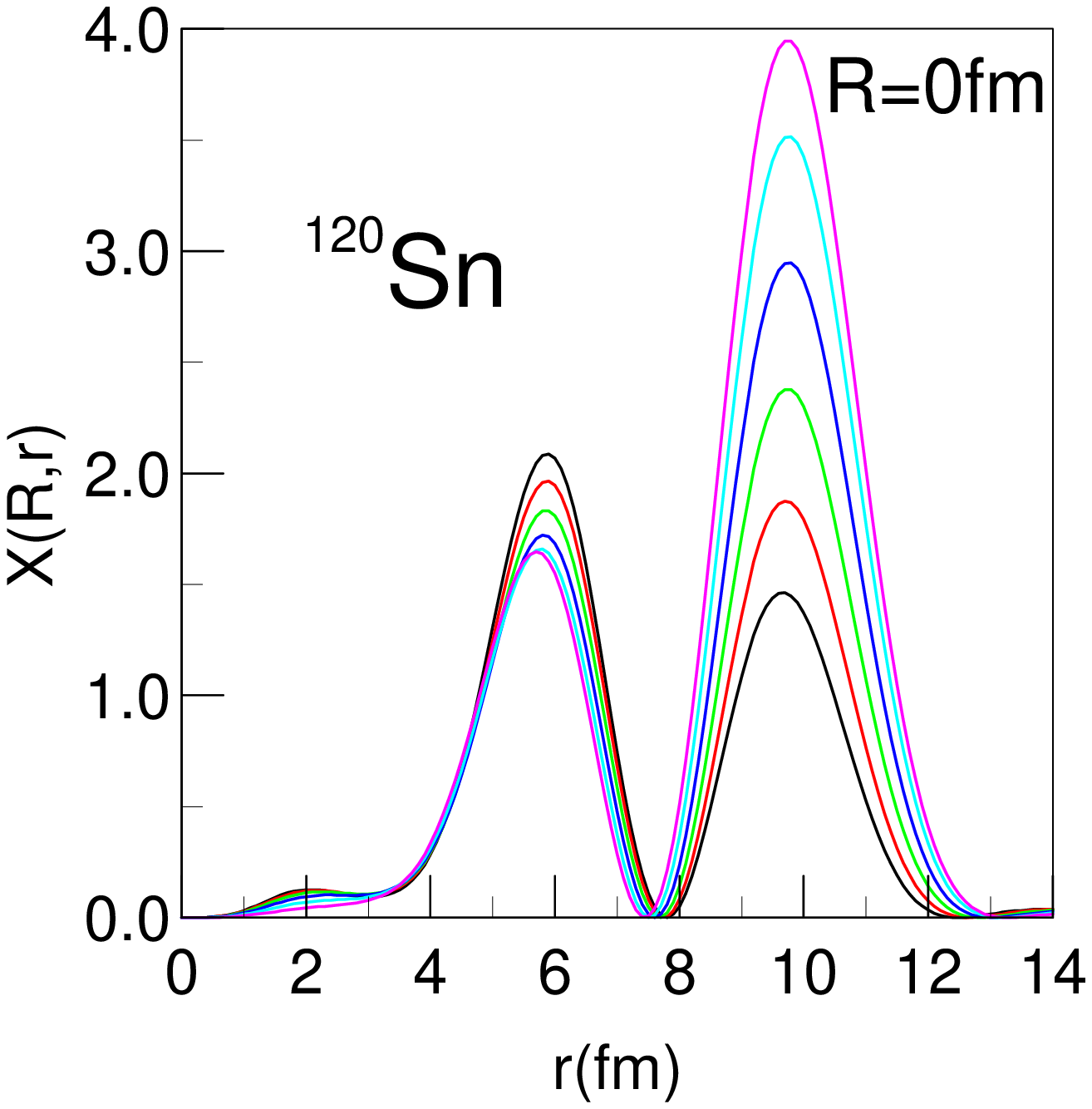}
\vspace{-1cm}
\includegraphics*[scale=0.45,angle=0]{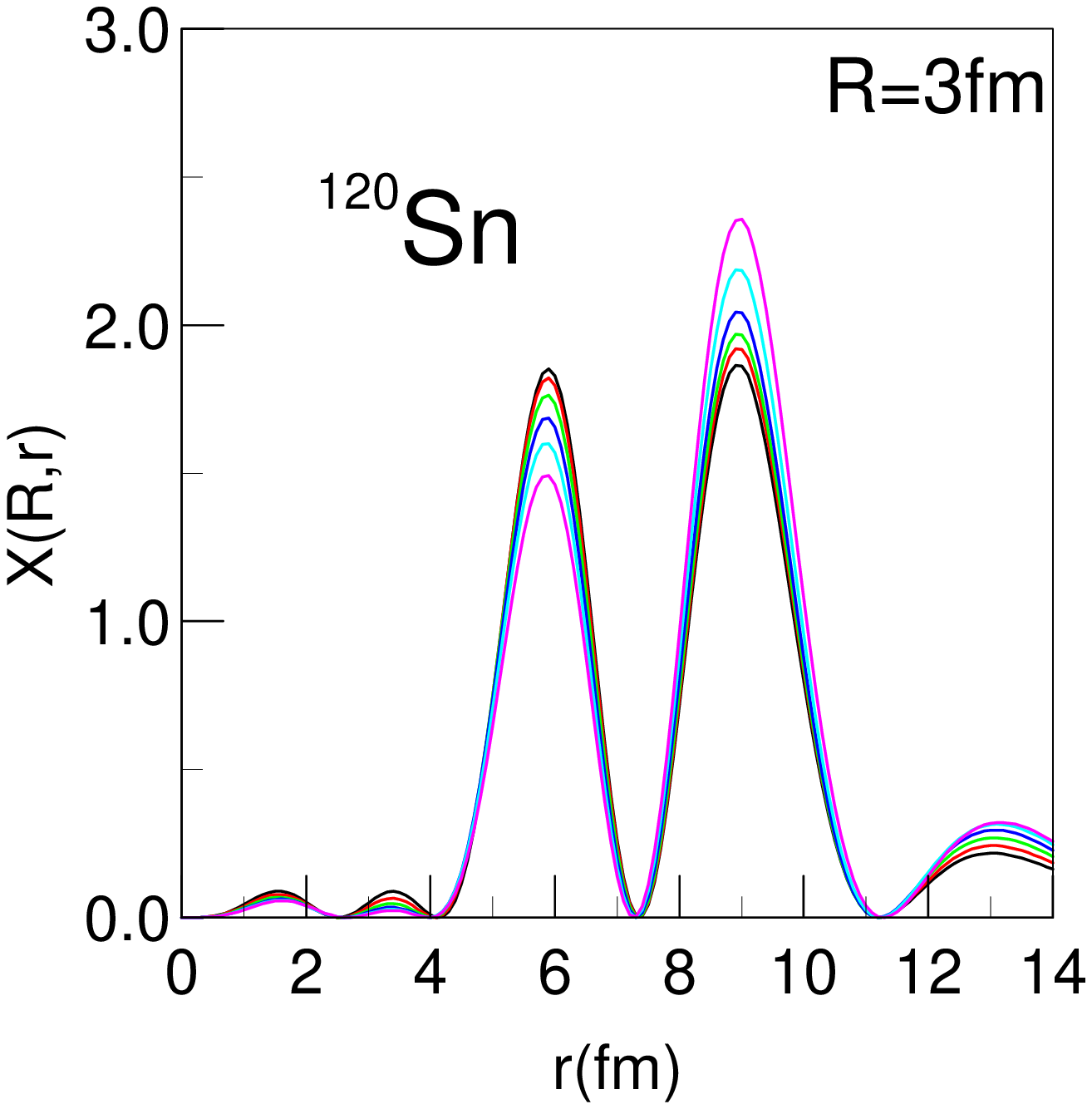}
\vspace{-1cm}
\includegraphics*[scale=0.45,angle=0]{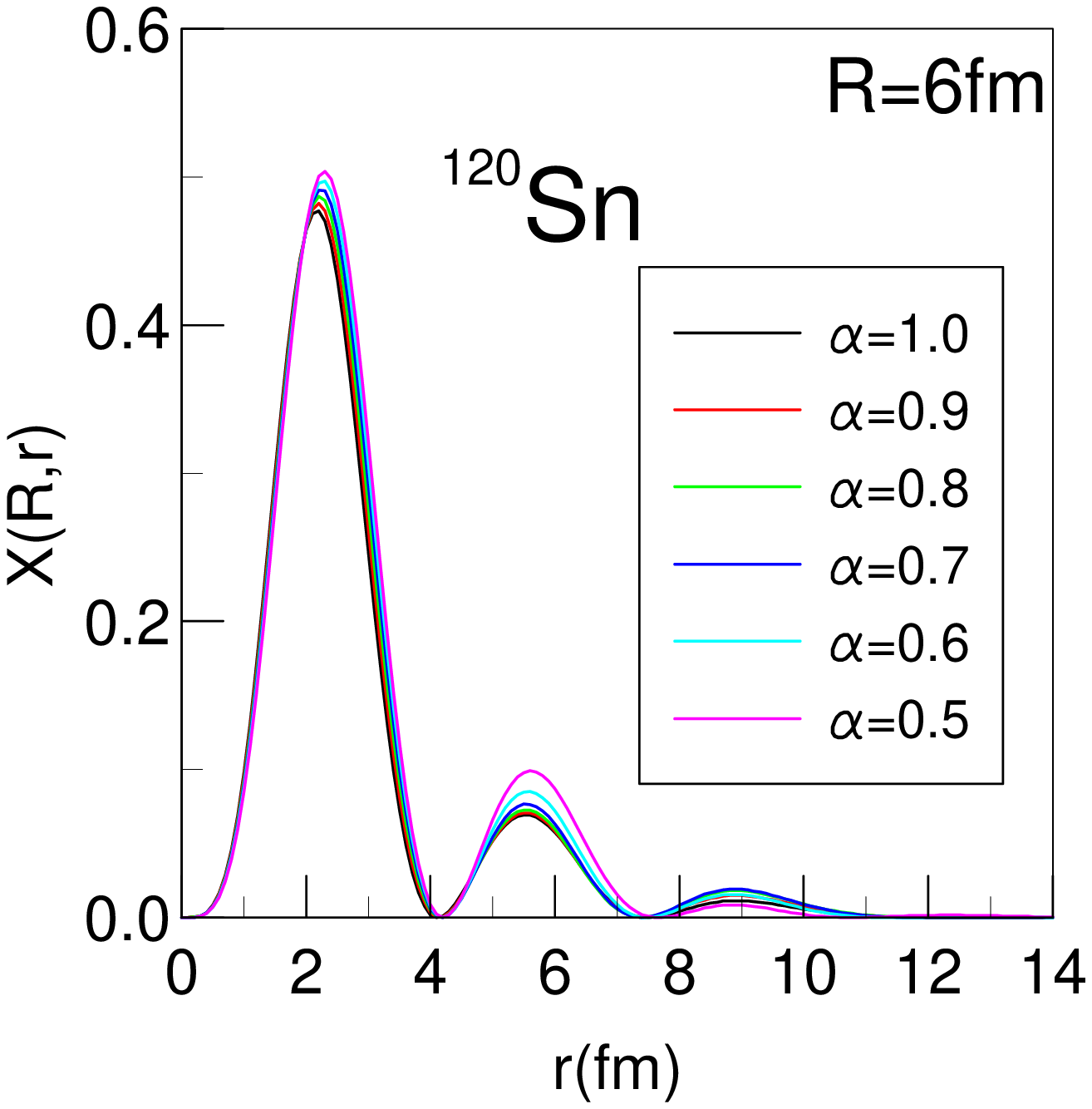}
\vspace{-1cm}
\includegraphics*[scale=0.45,angle=0]{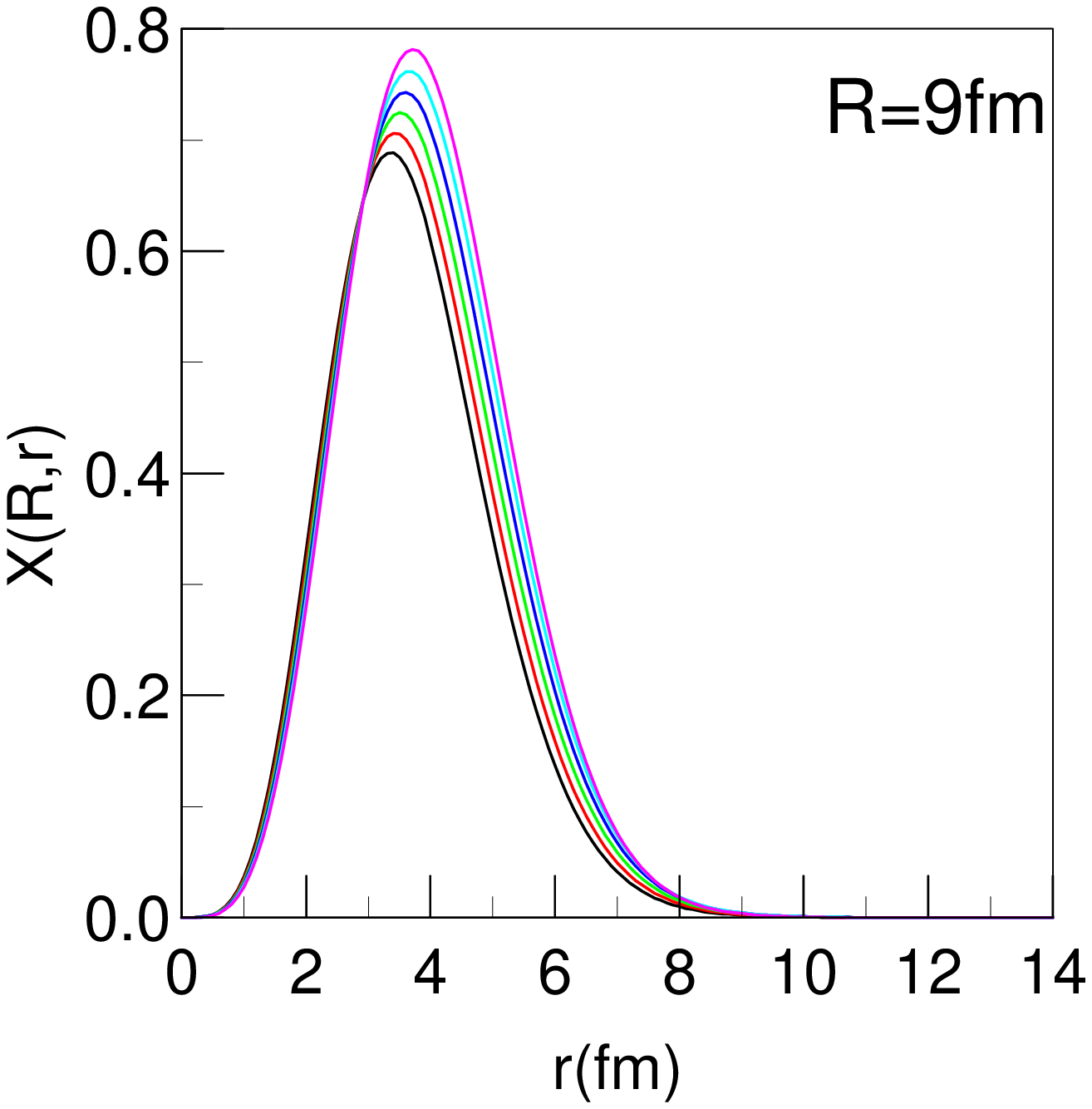}
\vspace{0.2cm}
\caption{(Color online) $X(R,r)$ for R=0, 3, 6 and 9 fm in the case of $^{120}Sn$.}
\label{new6b}
\end{figure}

$X(R,r)$ is presented in Fig.\ref{new6b} for four values of R namely, 0, 3, 6 and 9 fm corresponding to
the interior of the nucleus and the vicinity of the surface. The results are
displayed for various values of the pairing factor $\alpha$.

Except for $R=0$, $X(R,r)$ and hence $\xi(R)$ is not really sensitive to the strength of the
pairing interaction.
The large dependence of $X(R,r)$ on the pairing strength at $R=0$ comes from the comparatively large
parity mixing already mentioned in connection with Fig.\ref{new3}, which is negative and maximum in
absolute value for $r \simeq 10$ fm. Since the parity mixing tends to disappear
as the pairing strength
decreases, the height of the peak at $r=10$ fm increases. In contrast, for $R= 3,
6, 9$ fm the influence
of the parity mixing is very modest and the behaviour of $X(R,r)$ is determined essentially by
$\kappa_{e}^2$ and $\kappa_{o}^2$.
From $R \simeq 3$ fm to $R \simeq 6$ fm, one observes a sensitive reduction of
the magnitude of $X(R,r)$
leading to a lowering of the CL. In the vicinity of the surface ($R \ge 6$ fm), the oscillatory behaviour
of $X(R,r)$ disappears. Here, single particle wave functions have almost reached their exponential regime.
This explains why at $R \ge 6$ fm, $X(R,r)$ is characterized by only one major peak.
The width of this major peak is minimum at the nuclear surface.
Its broadening for $R=9$ fm explains the increase of the CL beyond the nuclear surface.

\begin{figure}
\includegraphics*[scale=0.45,angle=0]{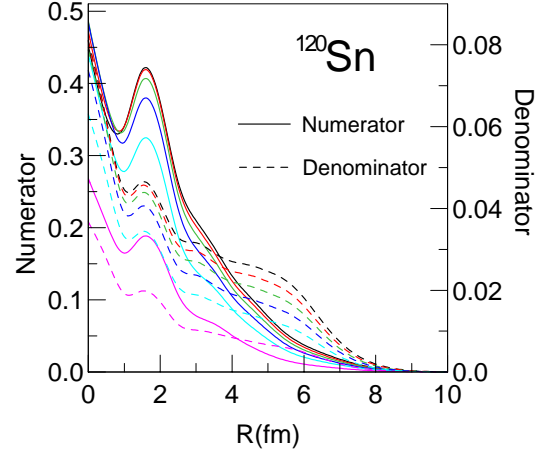}
\caption{Evolution of numerator and denominator of $\xi(R)$ for various values of $\alpha$, in the case of $^{120}Sn$.}
\label{new6}
\end{figure}

\begin{figure}
\includegraphics*[scale=0.45,angle=0]{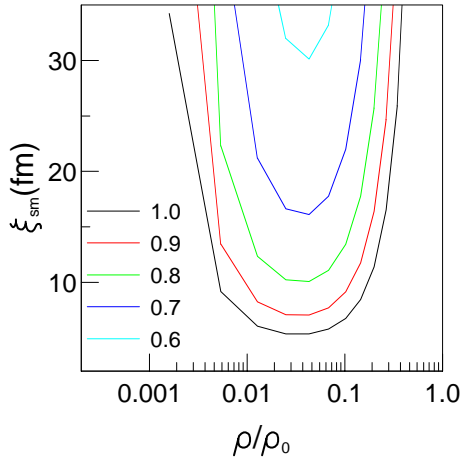}
\caption{(Color online) Coherence length calculated with different intensity of pairing strength
in symmetric nuclear matter.}
\label{xinm}
\end{figure}

A more global way to analyze the behavior of the CL is to 
consider directly the dependence on $R$ of the
numerator and the denominator of Eq.(\ref{eq20b}). This is 
shown in Fig.\ref{new6}.
One sees, that independently of the value of $\alpha$ (the color code is the same as for
Fig.\ref{intensity}), the denominator decreases faster than the numerator around $R=6$ fm and beyond.
This sudden change in the slope of the denominator is accountable for the minimum value of CL.

A similar analysis of the CL and of the influence of pairing correlations has been carried out in
infinite matter.
In Fig.\ref{xinm}, we show the CL in infinite symmetric nuclear matter as a function of the density $\rho$
normalized to its saturation value $\rho_{0}$, for the same $\alpha$ values as in the HFB calculations
for finite nuclei. In the nuclear matter case, we see that the CL depends very strongly on the pairing
intensity, whatever the density. For instance, the minimum value of CL increases a lot as pairing decreases.
This behavior can be understood from an approximate analytic evaluation of the CL in infinite nuclear
matter based on the definition Eq.(\ref{eq20b}) which differs only slightly from the usual Pippard expression
\cite{fetter} (see Appendix \ref{appendix1}):
\begin{equation}\label{eqcl}
\xi_{nm}= \frac{\hbar^2 k_F}{2 \sqrt{2}m^{*} |\Delta_{F}|}
\left(1+ \frac{a^2}{8} \left( 3b^2-12b+4 \right) + {\cal O}(a^3)\right)
\end{equation}
where $a=|\Delta_F|/|\epsilon_F|$, $b=k_F \Delta^{'}_{F}/\Delta_F$ with
$\Delta_F$ and $\Delta^{'}_{F}$ the pairing field and its derivative for the
Fermi momentum $k_F$. As discussed in Appendix \ref{appendix1}, the
correction terms in Eq.(\ref{eqcl}) are very small.

We see that the CL in infinite matter varies approximatively inversely
proportional to the gap at the Fermi surface.
This behavior is at variance with the results in finite nuclei and particularly where the CL shows
the minimum, see Fig.\ref{intensity}.
This clearly indicates that the behavior of the CL, in particular the
small value obtained in the surface of finite nuclei,
 is strongly influenced by the structure of the orbitals and that
pairing plays a secondary role.

\begin{figure}
\begin{center}
\includegraphics*[scale=0.45,angle=0]{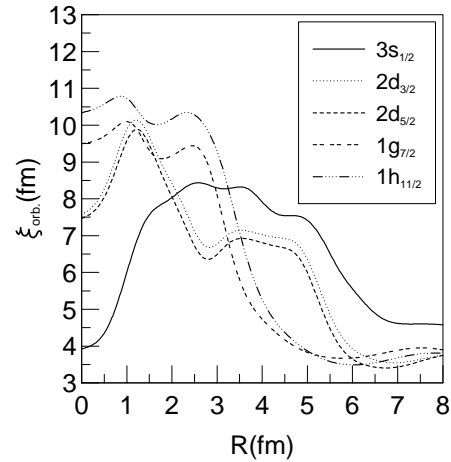}
\end{center}
\caption{ Coherence length for
Hartree-Fock single particle orbitals of the neutron valence shell of $^{120}$Sn.}
\label{orbital}
\end{figure}

In order to examine this question in more detail,
we show in Fig.\ref{orbital} the extension of completely uncorrelated pairs
made of Hartree-Fock neutron single particle wave functions.
We use a definition
of the extension of the pair size similar to the CL of Eq.(\ref{eq20b}) namely,
\begin{equation}
\mathrm{\xi_{orb}(R)}  = \frac
{\left(\int r^2 |A_{i}(\vec{R},\vec{r})|^2 d^3 r\right)^{1/2}}
{\left( \int |A_i(\vec{R},\vec{r})|^2 d^3 r\right)^{1/2}}
\end{equation}
The uncorrelated pair wave function $A_i(\vec{R},\vec{r})$ is defined as
\begin{equation}
\begin{array}{l}
\dspt A_i(\vec{R},\vec{r}) = \frac{1}{4 \pi} (2j_{i}+1) \sum_{n_{\alpha} n_{\beta}}
C_{n_{\alpha}}^{n_{i} l_{i} j_{i}} C_{n_{\beta}}^{n_{i} l_{i} j_{i}} \\
\dspt ~~~~~~\times \sum_{nNl} (-)^{l} \frac{(2l+1)^{1/2}}{2l_{i}}
 u_{nl} (r/ \sqrt{2}) u_{Nl} (\sqrt{2}R) \\
~~~~~~\times P_{l} (cos \theta) \langle nl Nl; 0 | n_{\alpha} l_{i} n_{\beta} l_{i} 0 \rangle
\end{array}
\end{equation}
where $C_{n_{\alpha}}^{n_{i} l_{i} j_{i}}$ is the component of the
$(n_{i} l_{i} j_{i})$ neutron single-particle orbital on the HO basis function
$(n_\alpha l_{i} j_{i})$. This equation is the same as Eq. (3) of
Ref.~\cite{ref1} with the matrix $\kappa^{l_ij_i}_{n_\alpha n_\beta}$ of the
pairing tensor replaced with the product of the two $C$
coefficients.\\
Since $\xi_{orb}(R)$ corresponds to two non-interacting
neutrons put into the same orbit and coupled to (L=0, S=0), it 
contains only the
correlations induced by the confinement of the single-particle wave functions.
As Fig.\ref{orbital} shows, $\xi_{orb}$ has a pattern rather
similar to the global CL displayed in Figs. \ref{fig8} and \ref{intensity},
except for the $3s_{1/2}$ orbital. Thus, provided this orbital is not
strongly populated, a change in the relative contributions of the
single-particle states in the pairing tensor,e.g., induced by varying the
intensity of pairing correlations, will not cause significant modifications in
the global CL. This result has also been found by Pastore \cite{pastore_p}.
From Fig.\ref{orbital} we see that (except for
$3s_{1/2}$), $\xi_{orb}(R)$ exhibits a minimum in the surface of the order of
$\simeq 3.5$ fm. This is indeed small
but still larger than the $2.3$ fm found with Eq.(\ref{eq20b}) for $\alpha=1$ (or $2.5$ fm 
for $\alpha=0.5$).
The reduction by about 30 percent from 3.5 fm to 2.3 fm of
the minimum of the CL is very likely due to the fact that even for very
small pairing some orbit mixing takes place (remember that the influence
of pairing is compensated in the ratio of numerator and denominator and
that the chemical potential becomes not necessarily locked to a definite
level but may stay in-between the levels). The cross terms of the wave
functions can be negative yielding a possible explanation of the effect.
Let us also point out that the CL implying only the even part
of the pairing tensor (or the odd one), see \cite{ref1}, is of the order
of $\sim $ 2.7 fm for $^{120}$Sn, see Fig.12. 
Therefore, there should exist a slight
influence of parity mixing in the CL calculated with the full $\kappa$.

Nonetheless, the above discussion clearly indicates that the small value of the
CL in the surface of finite nuclei is essentially due to the structure
of the single particle wave functions. Our conclusion is somewhat different 
from the one put forward in our early paper
\cite{ref1}. There, we had not explored the behavior of the CL as a function
of the pairing strength, which led us to conclude that the small size of Cooper
pairs stems from a local strong coupling pairing regime. However,
the other results and conclusions of Ref. \cite{ref1} still hold.

One may speculate about the reason for this radically different behavior of
the CL in nuclei and infinite nuclear matter. One issue which certainly can
be invoked, is that in macroscopic systems the number of single particle states
in an energy range of the order of the gap is huge whereas in nuclei we only
have a few states/MeV.
In order to examine more precisely such an effect,
let us consider, for convenience, the example of a spherical harmonic
oscillator potential.
We want to keep the essential finite size
effects but eliminate unessential shell effects. It is well known that this
can be achieved via the so-called Strutinsky smoothing. Single shells are
washed out and what remains is a continuum model with energy as
variable, instead of individual discrete quantum states.
We therefore can write for the pairing tensor in Wigner space:

\begin{equation}
\kappa(\vec{R},\vec{p}) = \int_E~ dE  \kappa(E) f(E;\vec{R},\vec{p})
\end{equation}

\noindent
where $\kappa(E)=u_Ev_E$ is the Strutinsky averaged
pairing tensor \cite{brack} and $f(E;\vec{R},\vec{p})$ is the
Strutinsky averaged Wigner transform of the density matrix on the energy
shell $E$ \cite{vinas1}. Integration of this quantity over energy
up to the Fermi level
yields the Strutinsky averaged density matrix in Wigner space. The latter
quantity is shown in Fig. 1 of Ref. \cite{schlomo}.

A particularity of the Strutinsky smoothed spherical harmonic oscillator is
that all quantities  depend on $\vec{R}$ and $\vec{p}$ only via the
classical Hamiltonian $H_{\mathrm{cl.}}(\vec{R},\vec{p})$. We see that the Wigner
transform of the density matrix is approximately
constant for energies below the Fermi energy and drops to zero within a width
of order $\hbar \omega$. The corresponding density matrix on the energy shell
can then be obtained from the quantity shown in Ref. \cite{schlomo} by differentiation with
respect to energy. We, therefore, deduce that
$f(E;H_{\mathrm{cl.}})$ is peaked around $E \sim H_{\mathrm{cl.}}$ with a width of order
$\hbar \omega$.

The above integral over $E$ in $\kappa(\vec{R},\vec{p})$
is, therefore, a convolution of two functions, one of width $\sim \Delta$
and the other one of width $\sim \hbar \omega$. As long as the gap is
smaller than $\hbar \omega$, the $\vec{R}$ and $\vec{p}$ behavior of $\kappa$
will be dominated by $f(E;H_{\mathrm{cl.}})$, i.e. by the oscillator wave functions.
This is what happens in finite nuclei.
On the contrary, in infinite matter or in LDA descriptions,
$f(E;\vec{R},\vec{p})$ is a $\delta$-function, $\delta(E-H_{\mathrm{cl.}})$, and then
the $\vec{R}$, $\vec{p}$ behavior of $\kappa (\vec{R},\vec{p})$ is entirely
determined by the width of $\kappa(E)$, i.e. by the intensity of pairing.

This interpretation qualitatively explains the very different
behaviors of the CL with respect to the magnitude of pairing in
finite nuclei and infinite matter.
It also explains why the value of the CL in the surface
of finite nuclei can be much smaller than the one calculated in infinite nuclear
matter at any density.

More quantitative investigations along this line are in 
preparation \cite{vinas2}.

\section{Conclusions} \label{sect3}

In this paper we have continued our study of the spatial properties of pairing
correlations in finite nuclei. We first generalized our previous work \cite{ref1}
to deformed nuclei and found that the spatial behaviour of pairing is rather
similar to the spherical case. This concerns, for instance, the
remarkably small value of the coherence length (CL) ($\simeq$ 2
fm) in the nuclear surface. More inside the nucleus, sometimes more
pronounced differences appear.
We then concentrated on the reason for this strong minimal extension of the
CL in the surface of nuclei. It was found that this feature is practically
independent of the intensity of pairing and even seems to survive in the
limit of very small pairing correlations. 
A detailed analysis of the quantities entering the definition of the 
CL indicates that,
in finite nuclei, the latter is mainly determined by the single
particle wave functions, i.e. by finite size effects.
This eliminates suggestions that the strong observed lowering of the CL 
in the nuclear surface
has something to do with especially strong pairing correlations in the 
surface \cite{ref1}, or in a surface layer,
i.e. with a 2D effect \cite{kanada}. A particular situation seems to 
prevail in the two neutron halo state of $^{11}$Li \cite{Hagino}.

We also made the same study in infinite nuclear matter. We found that in 
that case the CL
strongly depends on the gap and an approximate inverse proportionality 
between the gap and the CL could be established.
Concerning the reason why nuclei and infinite matter behave so
differently with respect to the CL, we put forward the fact that the number of
levels in the range of the gap value is huge in a macroscopic system
whereas there are only a handfull of levels in finite nuclei. 
In such situations the numerator and denominator in the definition of the coherence 
length have a similar dependence on pairing and its influence tends to cancel.
From this work, it appears that the CL may not be a good indicator of
the spatial structure of pairing correlations in the case of 
nuclei or of other finite systems with a weak coupling situation like certain
superconducting ultra small metallic grains \cite{metal}.
This fact should not make us forget that on other quantities 
pairing in nuclei can have, as well known, a strong effect. For example the pairing
tensor itself, as studied in this work, is very sensitive to parity mixing, 
see Fig.\ref{new3}, where a strong redistribution, i.e. a concentration of pairing 
strength along the c.o.m positions of the pairs takes place. Such a feature probably 
is responsible for the strong enhancement of pair transfer into superfluid nuclei 
\cite{oertzen}. This small extension of the pairing tensor in the relative coordinate 
may not only be present in the surface but also in the bulk, depending somewhat on 
the shell structure. However, on average a generic but moderate enhancement of pairing 
correlations (obtained with the D1S Gogny force) is present in the nuclear surface, 
see Fig.\ref{fig9}.
Further elaboration of these aspects will be given in a forthcoming paper
\cite{vinas2}.

\vskip 0.3cm

\noindent
{\bf Acknowledgement}
We thank A. Pastore for sending us clarifying informations and for
fruitfull discussions. We acknowledge also K. Hagino, Y. Kanada-Enyo,
M. Matsuo, A. Machiavelli, H. Sagawa and X. Vi\~{n}as for useful
discussions.  This work was partially supported by CNCSIS through
the grant IDEI nr. 270.

\appendix
\section{Neutron coherence length in infinite nuclear matter}\label{appendix1}
Introducing the Wigner transform
\begin{equation}\label{a1}
\kappa_W(\vec{R},\vec{k})=\int d^3r~
\kappa(\vec{R},\vec{r})   e^{i\vec{k}\vec{r}}
\end{equation}
of the HFB neutron pairing tensor $\kappa(\vec{R},\vec{r})$, the coherence
length (CL) defined by Eq. (\ref{eq20b}) can be rewritten
\begin{equation}\label{a2}
\xi(\vec{R})=\sqrt{\frac{\int d^3k
|\overrightarrow{\nabla}_{k}\kappa_W(\vec{R},\vec{k})|^2}
{\int d^3k |\kappa_W(\vec{R},\vec{k})|^2}}
\end{equation}
In infinite nuclear matter, $\kappa_W$ is independent of
$\vec{R}$, depends on $\vec{k}$ only through the length $k=|\vec{k}|$, and is
given by $\kappa_W(k)= \Delta(k)/2E(k)$,
where $\Delta(k)$ is the HFB neutron pairing field and
$E(k)=\sqrt{(e(k)-\mu)^2+\Delta(k)^2}$ the neutron quasiparticle energies with
$e(k)$ the single-neutron energies and $\mu$ the neutron chemical potential.
Substituting these expressions into (\ref{a2}) yields $\xi_{nm}=\sqrt{N/D}$ with
\begin{equation}\label{a3}
\begin{array}{l}
\dspt N\!\!=\!\!\int_0^\infty \!\!\!\!k^2 dk
\frac{(e(k)-\mu)^2
\left(\Delta'(k)(e(k)-\mu) - \Delta(k) e'(k)\right)^2}
{\left[(e(k)-\mu)^2+\Delta(k)^2\right]^3} \\
\dspt D \!\!=\int_0^\infty \!\!\!\! k^2 dk
\frac{\Delta(k)^2}{(e(k)-\mu)^2+\Delta(k)^2}
\end{array}
\end{equation}
The primed quantities are first derivatives. In order to be able to
express the integrals analytically, we introduce the three
following approximations
\begin{enumerate}
\item $\mu\simeq e(k_F)\equiv e_F$ where $k_F$ is the neutron Fermi momentum,
\item $e(k)\simeq \hbar^2k^2/(2m^*)$ where $m^*$ is the $k_F$-dependent
neutron effective mass,
\item in the usual situation of nuclear physics where the gap values are much smaller than the Fermi energy,
the functions under the two integrals (\ref{a3}) are sufficiently
peaked around $k=k_F$ so that one can take
$\Delta(k)\simeq \Delta(k_F)\equiv \Delta_F$ and
$\Delta'(k)\simeq \Delta'(k_F)\equiv \Delta'_F$.
\end{enumerate}
Using these assumptions and making the change of variables $k=xk_F$,
expressions (\ref{a3}) become
\begin{equation}\label{a4}
\begin{array}{l}
\dspt N\!\!=\!\!a^2 k_F \int_0^\infty \frac{x^2 (x^2-1)^2\left[b (x^2-1) - 2 x\right]^2}
{\left[(x^2-1)^2+a^2\right]^3}     dx  \\
\dspt D \!\!=a^2 k_F^3\int_0^\infty \frac{x^2}{(x^2-1)^2+a^2}  dx
\end{array}
\end{equation}
with $a=|\Delta_F/e_F|$, $b=k_f\Delta'_F/\Delta_F$. Assuming $a\neq 0$, the
integrals on the right hand sides of (\ref{a4}) can be calculated
analytically using contour integration in the complex plane and the method of
residues (more precisely, the integrand for $N$ can be broken into an
even function for which the integration range, as the one for $D$, can be
extended from $-\infty$ to $+\infty$ and integrated by the method of residues,
and an odd function which is easily integrated after the change of variable
$y=x^2$). One gets:
\begin{equation}\label{a5}
\begin{array}{rl}
&\dspt N\!\!=\!\!a^2 k_F \left[2\pi \left(a Y \sqrt{\frac{1+\sqrt{1+a^2}}{2}}
-X \sqrt{\frac{-1+\sqrt{1+a^2}}{2}}\right)
\right.\\ & \left.\dspt \hspace{15mm}-\frac{b}{4a} \left(\frac{3\pi}{2}
+3\cot^{-1}(a) - \frac{a}{1+a^2}\right)\rule{0mm}{6mm}\right] \\
&\dspt D \!\!=\frac{\pi}{2}a k_F^3 \sqrt{\frac{1+\sqrt{1+a^2}}{2}}
\end{array}
\end{equation}
where $X$ and $Y$ are functions of $a$ and $b$ given by

\begin{equation}\label{a6}
\begin{array}{rl}
&\dspt X=\frac{a^2b^2(4a^2+5)-2(1+a^2)(5a^2+2)}{64\,a^2(1+a^2)^2} \\
&\dspt Y=\frac{a^2b^2 (21a^4+35a^2+12) +4(1+a^2)(7a^2+4)}{128\,a^4(1+a^2)^2}
\end{array}
\end{equation}
Usually, $a$ is much smaller than one, even at small densities. Expanding the
above expressions around $a=0$, one obtains
\begin{equation}\label{a7}
\xi_{nm}\sim \frac{1}{a k_F\sqrt{2}}\left(1
+\frac{a^2}{8}(3b^2-12b+4) +{\cal O}(a^3) \right)
\end{equation}
With $a=|\Delta_F|/(\hbar^2 k_F^2/2m^*)$, the leading term yields
\begin{equation}\label{a8}
\xi_{nm}\sim \frac{1}{2\sqrt{2}}\frac{\hbar^2 k_F}{m^*|\Delta_F|}
\end{equation}
This expression is very close to the Pippard approximation of the CL
\cite{fetter}
\begin{equation}\label{a9}
\xi_{Pippard}= \frac{1}{\pi}\frac{\hbar^2 k_F}{m^*|\Delta_F|}
\end{equation}
the pre-factor being $1/2\sqrt{2}\sim 1/2.8$ instead of $1/\pi$. Usual values of
$a$ and $b$ show that the first correction term in (\ref{a7}) is very small.
For instance, in symmetric nuclear matter at one tenth the normal density, one
gets $k_F\simeq 0.6$ fm$^{-1}$, $a\simeq .2$ and $b\simeq .3$ with the Gogny
effective force, which yields $3. 10^{-3}$ for this term. The next terms can be
shown to be even smaller.
Moreover, numerical evaluations of the integrands in (\ref{a3}) for the
Gogny force show that the
three above approximations employed for deriving (\ref{a5}), in particular the third one,
are extremely well justified for densities ranging from zero to twice the normal density in
symmetric nuclear matter.

\end{document}